\documentclass[12pt,preprint]{aastex}
\usepackage{epsf}

\def\deg{\ifmmode {^\circ}\else {$^\circ$}\fi}
\def\degree{\ifmmode {^\circ}\else {$^\circ$}\fi}
\def\mum{\ifmmode {\rm \,\mu {\rm m}}\else $\rm \,\mu {\rm m}$\fi}
\def\arcsec{\ifmmode ^{\prime \prime}\else $^{\prime \prime}$\fi}

\def\inch{\ifmmode ^{\prime \prime}\else $^{\prime \prime}$\fi}
\def\msunyr{\ifmmode {M_{\odot}~{\rm yr^{-1}}}\else $M_{\odot}~{\rm yr^{-1}}$\fi}
\def\msun{\ifmmode {M_{\odot}}\else $M_{\odot}$\fi}
\def\rsun{\ifmmode {R_{\odot}}\else $R_{\odot}$\fi}
\def\lsun{\ifmmode {L_{\odot}}\else $L_{\odot}$\fi}
\def\mstar{\ifmmode {M_{\star}}\else $M_{\star}$\fi}
\def\rstar{\ifmmode {R_{\star}}\else $R_{\star}$\fi}
\def\tstar{\ifmmode {T_{\star}}\else $T_{\star}$\fi}
\def\lstar{\ifmmode {L_{\star}}\else $L_{\star}$\fi}
\def\md{\ifmmode {M_d}\else $M_d$\fi}
\def\ld{\ifmmode {L_d}\else $L_d$\fi}
\def\ad{\ifmmode A_d\else $A_d$\fi}
\def\ldlstar{\ifmmode L_d / L_\star\else $L_d / L_{\star}$\fi}
\def\rearth{\ifmmode {\rm R_{\oplus}}\else $\rm R_{\oplus}$\fi}
\def\mearth{\ifmmode {\rm M_{\oplus}}\else $\rm M_{\oplus}$\fi}
\def\qdstar{\ifmmode Q_D^\star\else $Q_D^\star$\fi}
\def\vsqd{\ifmmode v^2 / Q_D^\star\else $v^2 / Q_D^\star$\fi}
\def\kms{\ifmmode {\rm km~s^{-1}}\else $\rm km~s^{-1}$\fi}
\def\ms{\ifmmode {\rm m~s^{-1}}\else $\rm m~s^{-1}$\fi}
\def\vrel{\ifmmode v_{rel}\else $v_{rel}$\fi}
\def\mesc{\ifmmode m_{esc}\else $m_{esc}$\fi}
\def\rmin{\ifmmode r_{min}\else $r_{min}$\fi}
\def\rmax{\ifmmode r_{max}\else $r_{max}$\fi}
\def\xmax{\ifmmode x_{max}\else $x_{max}$\fi}
\def\mmin{\ifmmode m_{min}\else $m_{min}$\fi}
\def\mmax{\ifmmode m_{max}\else $m_{max}$\fi}
\def\xcc{\ifmmode x_{cc}\else $x_{cc}$\fi}
\def\rmind{\ifmmode r_{min,d}\else $r_{min,d}$\fi}
\def\rmaxd{\ifmmode r_{max,d}\else $r_{max,d}$\fi}
\def\mmaxd{\ifmmode m_{max,d}\else $m_{max,d}$\fi}
\def\vrad{\ifmmode v_{rad}\else $v_{rad}$\fi}
\def\qz{\ifmmode q_{0}\else $q_{0}$\fi}
\def\qi{\ifmmode q_{i}\else $q_{i}$\fi}
\def\ql{\ifmmode q_{l}\else $q_{l}$\fi}
\def\qs{\ifmmode q_{s}\else $q_{s}$\fi}
\def\rbrk{\ifmmode r_{brk}\else $r_{brk}$\fi}
\def\rdamp{\ifmmode r_{damp}\else $r_{damp}$\fi}
\def\rin{\ifmmode r_{in}\else $r_{in}$\fi}
\def\rout{\ifmmode r_{out}\else $r_{out}$\fi}
\def\tin{\ifmmode t_{in}\else $t_{in}$\fi}
\def\tout{\ifmmode t_{out}\else $t_{out}$\fi}
\def\ain{\ifmmode a_{in}\else $a_{in}$\fi}
\def\aout{\ifmmode a_{out}\else $a_{out}$\fi}
\def\r0{\ifmmode r_{0}\else $r_{0}$\fi}
\def\R0{\ifmmode R_{0}\else $R_{0}$\fi}
\def\m0{\ifmmode m_{0}\else $m_{0}$\fi}
\def\M0{\ifmmode M_{0}\else $M_{0}$\fi}
\def\xm{\ifmmode x_{m}\else $x_{m}$\fi}
\def\sigz{\ifmmode \Sigma_0\else $\Sigma_0$\fi}
\def\gyr{\ifmmode {\rm g~yr^{-1}}\else ${\rm g~yr^{-1}}$\fi}
\def\cms{\ifmmode {\rm cm~s^{-1}}\else ${\rm cm~s^{-1}}$\fi}
\def\gcms{\ifmmode {\rm g~cm^{-2}}\else $\rm g~cm^{-2}$\fi}
\def\gcmc{\ifmmode {\rm g~cm^{-3}}\else $\rm g~cm^{-3}$\fi}
\def\atil{\ifmmode {\tilde{a}}\else $\tilde{a}$\fi}
\def\ttil{\ifmmode {\tilde{t}}\else $\tilde{t}$\fi}
\def\sqrttt{\ifmmode {\tilde{t}^{1/2}}\else $\tilde{t}^{1/2}$\fi}
\def\akari{{\it AKARI}}
\def\iso{{\it ISO}}
\def\iras{{\it IRAS}}

\def\herschel{{\it Herschel}}
\def\spitz{{\it Spitzer}}

\def\wise{{\it WISE}}
\def\orch{{\it Orchestra}}

\begin{document}

\title{Variations on Debris Disks IV. An Improved 
Analytical Model for Collisional Cascades}
\vskip 7ex
\author{Scott J. Kenyon}
\affil{Smithsonian Astrophysical Observatory,
60 Garden Street, Cambridge, MA 02138} 
\email{e-mail: skenyon@cfa.harvard.edu}

\author{Benjamin C. Bromley}
\affil{Department of Physics, University of Utah, 
201 JFB, Salt Lake City, UT 84112} 
\email{e-mail: bromley@physics.utah.edu}
%
%

\begin{abstract}

We derive a new analytical model for the evolution of a collisional
cascade in a thin annulus around a single central star.  In this model,
\rmax\ the size of the largest object changes with time, 
$\rmax\ \propto t^{-\gamma}$, with $\gamma \approx$ 0.1--0.2.
Compared to standard models where \rmax\ is constant in time,
this evolution results in a more rapid decline of \md\ the total mass
of solids in the annulus and \ld\ the luminosity of small particles in
the annulus: $M_d \propto t^{-(\gamma + 1)}$ and 
$L_d \propto t^{-(\gamma/2 + 1)}$. 
We demonstrate that the analytical model provides an excellent match to 
a comprehensive suite of numerical coagulation simulations for annuli at 
1~AU and at 25~AU. If the evolution of real debris disks follows the 
predictions of the analytical or numerical models, the observed 
luminosities for evolved stars require up to a factor of two more mass 
than predicted by previous analytical models.

\end{abstract}


\section{INTRODUCTION}
\label{sec: intro}

For over three decades, observations from \iras, \iso, \akari, \spitz,
\herschel, and \wise\ have revealed infrared excess emission from optically 
thin rings and disks of small solid particles surrounding hundreds of main 
sequence stars \citep[e.g.,][]{bac1993,wyatt2008,matthews2014,kuchner2016}.
Together with occasional direct images, the data suggest typical dust 
temperatures, 30--300~K, and luminosities, $\sim 10^{-5}$--$10^{-2}$,
relative to the central star. Although young A-type stars have the 
highest frequency of these `debris disks,' disks around young FGK 
stars are also common.  Binary systems are almost as likely to harbor 
debris disks as apparently single stars \citep{trill2007,stauffer2010,
kennedy2012,rodr2012,rodr2015}. Among all stars, the frequency of
debris disks declines roughly linearly with stellar age
\citep[e.g.,][]{rieke2005,currie2008,carp2009a,carp2009b,kenn2013a}.

Interpreting observations of debris disks requires a physical model which 
predicts observable properties of the solid particles as a function 
of stellar spectral type and age. The currently most popular model 
involves a collisional cascade within material left over from planet 
formation \citep[e.g.,][]{aum1984,bac1993,wyatt2002,kb2002b,dom2003,
krivov2006,wyatt2008,matthews2014}.
In this picture, planets excite the orbits of leftover planetesimals.
Destructive collisions among the planetesimals produce small dust 
grains which scatter and absorb/reradiate light from the central star.
As radiation pressure removes the smallest grains, ongoing collisions
replenish the debris. Over time, gradual depletion of the solid 
reservoir reduces the disk luminosity; the debris disk slowly fades 
from view.

Although analytical and numerical calculations of debris disks successfully
account for many observations, the models have a major inconsistency. In
analytical models, the radius of the largest objects undergoing destructive 
collisions (\rmax) is fixed in time \citep{wyatt2002,dom2003,wyatt2007a,
wyatt2007b,koba2010a,wyatt2011}. 
At late times, the disk mass \md\ and luminosity \ld\ in a thin annulus 
then decline linearly with time, $L_d, M_d \propto t^{-n}$ with 
$n \approx 1$. In numerical simulations, collisions gradually reduce the 
size of the largest object; \rmax\ then declines with time \citep[e.g.,]
[]{kb2002b,kb2008,kb2016a}. As a result, \ld\ and \md\ decline somewhat 
more rapidly ($n \approx$ 1.1--1.2) than predicted by the analytical model.

To reconcile the two approaches, we develop an analytical theory for the
decline of \rmax\ with time. Combining our result with the standard theory
for the decline of the disk mass leads to a self-consistent picture for
the long-term evolution of \rmax, \md, and \ld\ which generally matches 
the results of numerical simulations. The new theory should enable more 
robust comparisons of models with observations of debris disks.

After briefly summarizing existing theory, we formulate and solve an
analytical model for the evolution of \rmax\ in \S\ref{sec: theory}.
In addition to matching current theory when \rmax\ is constant, the 
model predicts how the decline of \rmax\ with time depends on the 
physical properties of the solids in the disk. The analytical
solutions for \rmax\ agree remarkably well results from a suite of 
numerical simulations (\S\ref{sec: comps}). In \S\ref{sec: summary},
we conclude with a brief summary.

\section{EXPANDED ANALYTIC MODEL}
\label{sec: theory}

In the standard analytic model for collisional cascades, solid particles with radius $r$, 
mass $m$, and mass density $\rho$ orbit with eccentricity $e$ and inclination $i$ inside
a cylindrical annulus with width $\delta a$ centered at distance $a$ from a central star 
with mass \mstar\ and luminosity \lstar.  For particles smaller than some maximum size 
\rmax\ (mass, \mmax), all collisions are destructive.  Among particles ejected in a
collision, radiation pressure removes those smaller than some minimum size \rmin\ (mass,
\mmin). This loss of material leads to a gradual reduction in the total mass $M_d$ with 
time.  If the swarm of particles has a size distribution $N(r)$, integrating the collision 
rate over all sizes $r \le \rmax$ yields the time evolution of the total mass, $M_d(t)$ 
\citep[e.g.,][]{dohn1969,hellyer1970,will1994,obrien2003,koba2010a,wyatt2011,kb2016a}.  

To expand the analytical theory to include a changing \rmax, we separate collisions into
cratering and catastrophic regimes \citep[see also][and references therein]{krivov2006,
koba2010a,wyatt2011}.  For a collision between two particles with masses $m_1$ and $m_2$ 
($m_2 \le m_1$) and radii $r_1$ and $r_2$ ($r_2 \le r_1$), catastrophic collisions result 
in a cloud of debris with a mass similar to the combined mass of the colliding particles 
and particle sizes much smaller than $r_1$. In cratering outcomes, the ejected mass is 
often larger than $m_2$ but significantly smaller than $m_1$; thus, $m_1$ loses mass. 
Our goal is to derive an analytical prescription for the change in \rmax\ from cratering.

We begin our derivation with the collision time $t_0$. For a swarm of identical solid 
particles with radius \rmax\ \citep{wyatt2002,dom2003,wyatt2008,koba2010a,wyatt2011,
kb2016a}:
\begin{equation}
t_0 = { r_0 \rho P \over 12 \pi \Sigma_0} ~ ,
\label{eq: t0}
\end{equation}
where $r_0$ is the initial radius of the largest particles in the swarm, $P$ is the 
orbital period, $\Sigma_0 = M_0 / 2 \pi a \delta a$ is the initial surface density 
of solids, and $M_0$ is the initial mass of the swarm. By construction, collisions 
among these largest particles are catastrophic.

To simplify comparisons with previously published expressions for $t_0$ 
\citep[e.g.,][]{wyatt2002,dom2003,krivov2005,krivov2006,koba2010a}, we express 
$t_0$ in terms of the initial cross-sectional area of the swarm, $A_0$. Adopting 
$M_0 / A_0 = 4 \rho r_0 / 3$, $t_0 = 2 \pi a \delta a P / A_0$. In this form,
the collision time depends only on the geometry of the annulus, the orbital period, 
and the cross-sectional area of the swarm.

In an ensemble of mono-disperse objects with radius \rmax\ and total mass $M_d$, 
the instantaneous mass loss rate is $\dot{M} = -M_d / t_{max}$, where $t_{max}$ 
is the collision time.  When the swarm contains particles with radii smaller than 
\rmax, the collision time depends on the relative number of cratering and 
catastrophic collisions and the way these collisions re-distribute mass through 
the swarm. To quantify this process, we set $\dot{M} = -M_d / \alpha t_{max}$.
Initially, $t_{max} = t_0$; as the swarm evolves, \rmax\ and $M_d$ grow smaller.  
Setting $t_{max} = (\rmax/r_0) (M_0 / M_d) t_0$ allows us to relate the evolving 
collision time to changes in $M_d$ and \rmax. Smaller $r_0$ ($M_d$) results in 
shorter (longer) collision times.

These definitions yield a simple differential equation for $M_d(t)$ that depends on
the initial state of the system and the two unknowns \rmax\ and $M_d$:
\begin{equation}
\dot{M}_d = - \left ( { M_d^2 \over \alpha M_0 t_0 } \right ) \left ( {r_0 \over \rmax } \right ) ~ .
\label{eq: mdisk}
\end{equation}
With $\rmax \le r_0$, $M_d$ declines more rapidly with time compared to models 
with constant \rmax.

Deriving $\alpha$ requires a collision model. Following methods pioneered by
\citet{saf1969}, the rate particles with radius $r_1$ experience collisions
with all particles with radius $r_2 \le r_1$ is $n_2 \sigma v$, where $n_2$ 
is the number density of smaller particles, $\sigma$ is the cross-section, 
and $v$ is the collision velocity.  To express this rate in terms of the 
properties of the swarm, we adopt the formalism developed for our numerical 
simulations of planet formation \citep[e.g.,][and references therein]{kl1998,
kb2002a,kb2004a,kb2008,kb2012,kb2016a}.  Specifically, 
\begin{equation}
{dN_1 \over dt} = { \epsilon N_1 N_2 (r_1 + r_2)^2 \Omega \over 4 a \delta a } ~ ,
\end{equation}
where $N_1$ ($N_2$) is the number of particles with radius $r_1$ ($r_2$), 
$\Omega = 2 \pi / P$ 
is the angular velocity of particles orbiting the central star, and 
$\epsilon \simeq$ 1.044 is a factor that includes geometric factors in 
the cross-section, the distribution of particle velocities, and the ratio 
$i/e$ = 0.5 for the swarm. For this derivation, we assume the gravitational
focusing factor is unity.

Collision outcomes depend on the ratio of the collision energy $Q_c$ to the 
binding energy \qdstar. Here, we assume \qdstar\ is independent of particle
size. After a collision, the mass of the combined object is $m = m_1 + m_2 - m_e$ 
where $m_e$ is the mass that escapes as debris. In our approach, 
$Q_c = m_1 m_2 v^2 / 2 (m_1 + m_2)^2$ and $m_e = 0.5 (m_1 + m_2) (Q_c / \qdstar)^{b_d}$,
where $b_d$ is a constant of order unity.  
Setting $ x = r_2 / r_1$,
\begin{equation}
m_e = \left ( { m_2 \over 4 (1 + x^3) } \right ) \left ( { v^2 \over \qdstar } \right )^{b_d} ~ .
\label{eq: me}
\end{equation}
Depending on $v^2 / \qdstar$, the 
ejected mass ranges from zero to the combined mass $m_1 + m_2$.  For equal mass 
particles ($x$ = 1), catastrophic collisions eject half of the combined mass 
when $v^2 / \qdstar$ = 8.

The fate of the ejected mass depends on the size distribution. Although 
numerical calculations provide some guidance on the ejecta at large sizes 
\citep[e.g.,][]{benz1999,durda2004,durda2007,lein2008,lein2009,morby2009b,
lein2012}, there is little information on small sizes
\citep[e.g.,][]{krijt2014}. 
For simplicity, we adopt a standard power law $N_e(r) \propto r^{-3.5}$ 
\citep[see also][and references therein]{koba2010a,weiden2010b,wyatt2011},
where the size of the largest object in the debris is 
\begin{equation}
m_l = 0.2 \left ( {v^2 \over \qdstar } \right )^{-b_l} m_e ~ 
\label{eq: mlg}
\end{equation}
and $b_l$ is another constant of order unity.  If radiation pressure 
removes all particles with mass $m \le \mmin$, the amount of mass lost 
in each collision is then $ m_e (\mmin / m_l)^{1/2}$. 

With expressions for $dN_1 / dt$, $m_e$, and $m_l$, we can derive 
$\dot{M}_d$ by integrating the mass loss rate for a single collision 
over $r_2$ and $r_1$:
\begin{equation}
\dot{M}_d = - \int \int \delta_{12} m_e ~ \left ( { \mmin \over m_l } \right )^{1/2} ~ { dN_1 \over dt } ~ dr_1 dr_2 ~ ,
\label{eq: dmdt}
\end{equation}
where $\delta_{12}$ is a factor which prevents double-counting of collisions 
among identical particles.
Accomplishing this task requires a simple numerical integration. We divide 
particles into a set of $N_b$ logarithmic mass bins ranging in size from 
\rmin\ to \rmax\ with a ratio $\delta_r = r_{i+1} / r_i$ between bins. For 
an adopted size distribution $N(r)$, our algorithm establishes the mass in 
each bin and then integrates over the bins to infer the mass loss rate. For 
any set of initial conditions, 
\begin{equation}
\alpha = - \left ( { M_0 \over t_0 } \right ) \dot{M}_d^{-1} ~ .
\label{eq: alpha} 
\end{equation}
Experiments with different $\delta r$ suggest that the integrals converge
to better than 0.1\% with 2048--4096 mass bins between \rmin\ = 1~\mum\ and
\rmax\ = 100~km.

For this analysis, we consider two initial size distributions. In the
simplest case, $N(r) = N_0 r^{-3.5}$ where $N_0$ is a constant which 
sets the total mass of the swarm, $M_0 = (8 \pi \rho / 3) N_0 \rmax^{1/2}$
when $\rmax \gg \rmin$. In an equilibrium collisional cascade, however,
the size distribution develops a wavy pattern superimposed on the simple
power law \citep{campo1994a,obrien2003,wyatt2011}. For cascades where
catastrophic collisions dominate, \citet{kb2016a} derive a recursive 
solution for the equilibrium size distribution from a formalism developed 
by \citet{wyatt2011}. \citet{kb2016a} also show that numerical solutions 
to collisional cascades which include cratering yield size distributions 
reasonably close to the analytical result.

To compare solutions for $\alpha$ with different initial size distributions,
we consider debris in an annulus with $\Sigma_0$ = 10~\gcms, $a$ = 1~AU, 
$\delta a$ = 0.2~AU. Particles have sizes ranging from \rmin\ = 1~\mum\ to
\rmax\ = 100~km and mass density $\rho$ = 3~\gcmc. We also set $b_d$ = 1 and 
$b_l$ = 1. For these starting conditions, $t_0 \simeq 7.96 \times 10^4$~yr. 
With the power law initial size distribution, we derive $\alpha$ for 
$\vsqd\ \ge 1$. In our formalism, we construct equilibrium size distributions
only in systems where collisions between equal mass objects are catastrophic,
e.g., $\vsqd\ \ge$ 8.  Thus, we do not infer $\alpha$ for systems with 
\vsqd\ = 1--8 and the equilibrium size distribution. For either initial size 
distribution, the derived $\alpha$ is somewhat sensitive to $b_d$ and $b_l$
but is independent of $a$, $\delta a$, $\Sigma_0$, \rmin, \rmax, and $\rho$.

\begin{figure} 
\includegraphics[width=6.5in]{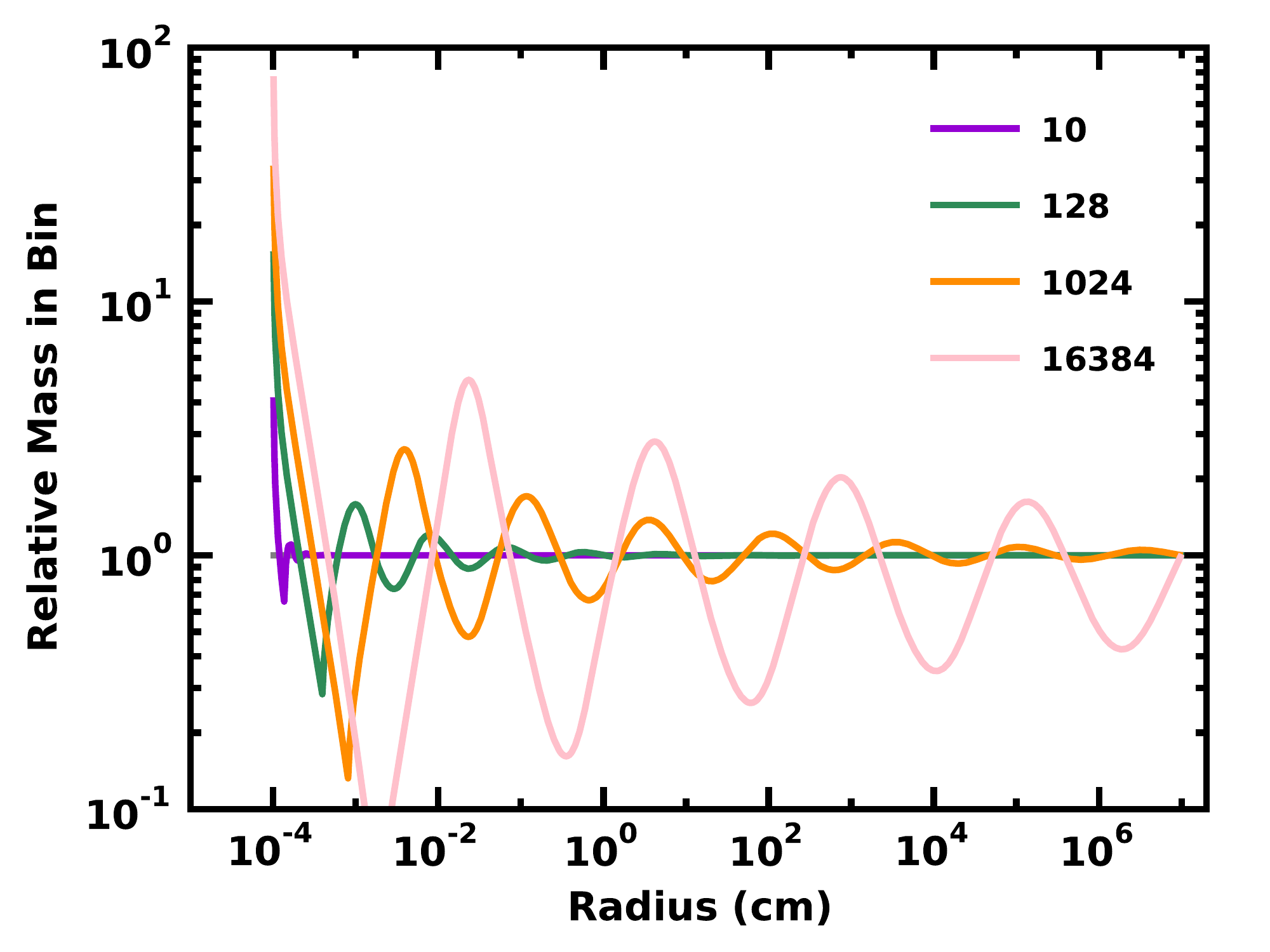}
\vskip 3ex
\caption{
Relative mass distributions for equilibrium models of collisional
cascades. The legend indicates $v^2/\qdstar$ for each curve.  Systems 
with larger $v^2 / \qdstar$ have more wavy mass distributions.
\label{fig: equil}
}
\end{figure}

Fig.~\ref{fig: equil} compares the relative mass distributions for 
equilibrium solutions with different values of $v^2 / \qdstar$. In 
systems with the simple power law ($N(r) \propto r^{-3.5}$), the
relative mass distribution follows a straight horizontal line. For
equilibrium mass distributions, the lack of grains with $r \le \rmin$
prevents collisional disruption of particles with $ r \approx$ 
1--3~\rmin\ and produces an excess of these objects \citep{campo1994a,
obrien2003,wyatt2011,kb2016a}. Similarly, the excess of particles just
larger than \rmin\ produces a deficit of particles with $r \approx$
10~\rmin. At small $\vsqd$, the waviness in the relative mass
distribution is minimal and confined to particle sizes $r \lesssim$ 
10--30~\rmin.  As the adopted $\vsqd$ grows, the relative mass 
distribution becomes wavier and wavier at larger and larger sizes.

Along with dramatic changes in waviness as a function of \vsqd, these 
size distributions have very different ratios of the cross-sectional 
area (\ad) to the total mass of the swarm (\md). In a standard power-law
size distribution, $N \propto r^{-3.5}$, with \rmin\ = 1~\mum\ and 
\rmax\ = 100~km, $\md/\ad \approx$ $12.65 (r_{max} / {\rm 1~km})^{1/2}$. 
In wavy size distributions with \vsqd\ = 8, $\ad/\md$ is identical to 
the power-law ratio.  The derived $\md / \ad$ slowly drops with 
increasing \vsqd, falling by a factor of roughly 3 (10) when 
\vsqd\ = $10^3$ ($10^5 - 10^6$).  For \vsqd\ $\le 10^3$, decline in 
$\md / \ad$ is fairly independent of \rmax. At larger \vsqd, the 
amount of waviness and $\md / \ad$ are more sensitive to \rmax.

With $L_d \propto \ad$, systems with the equilibrium size distribution 
and \vsqd $\ge$ 10 require less mass to produce the same infrared excess.
This mass monotonically decreases with increasing \vsqd.

\begin{figure} 
\includegraphics[width=6.5in]{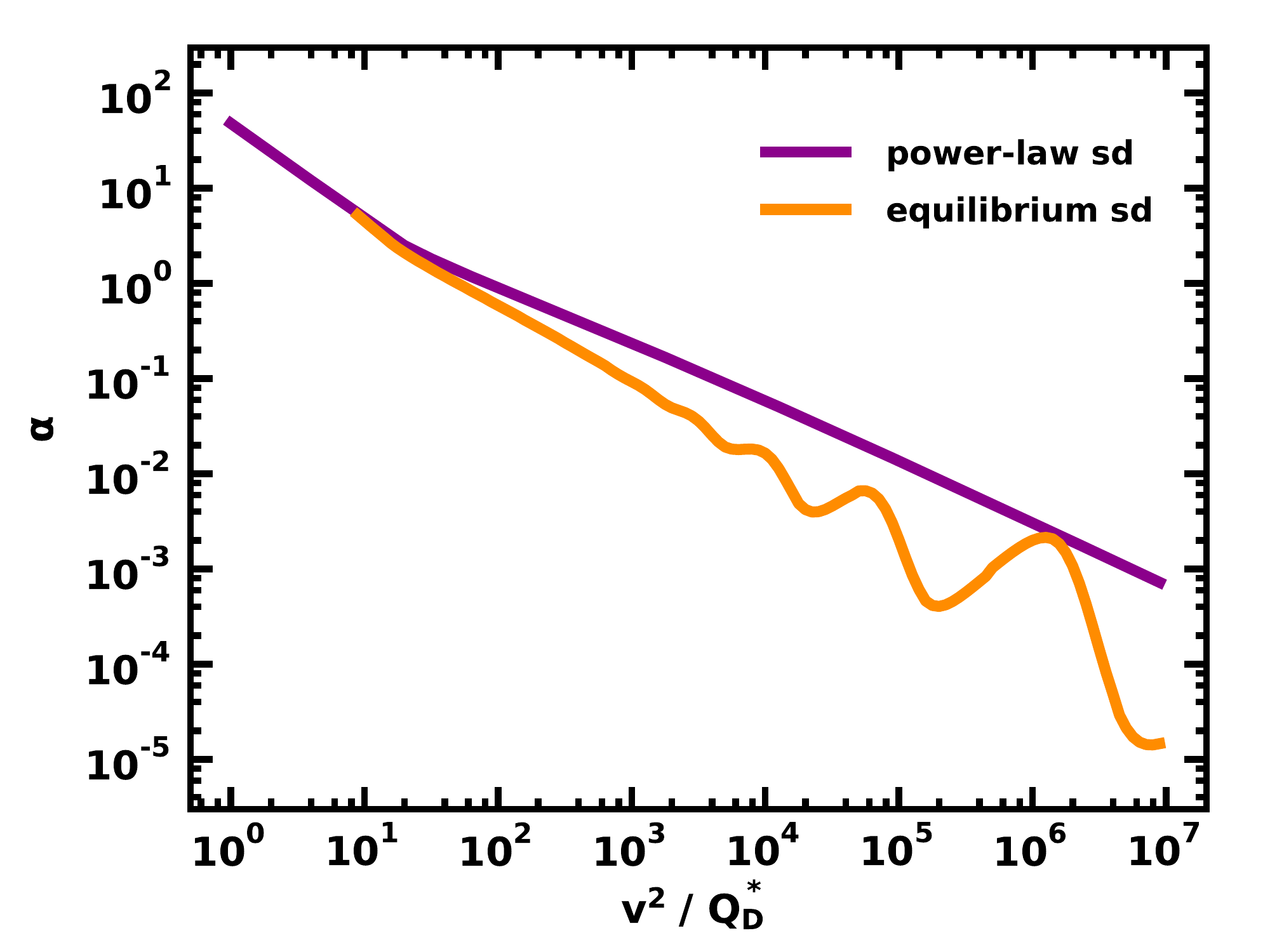}
\vskip 3ex
\caption{
Collision time parameter $\alpha$ for power-law (magenta symbols) 
and equilibrium (orange symbols) size distributions. With
$t_c = \alpha t_0$, systems with larger $v^2 / \qdstar$ have
shorter collision times. For equilibrium models, extremely wavy 
size distributions at large $v^2 / \qdstar$ yield wavy behavior 
in $\alpha$.
\label{fig: alpha1}
}
\end{figure}

Fig.~\ref{fig: alpha1} illustrates the impact of the adopted size
distribution on $\alpha$ for a broad range of \vsqd.  When 
$\vsqd\ \le 8$, most collisions eject little mass from the combined 
object. With $\dot{M}_d$ small, $t_c$ is larger than $t_0$. As
\vsqd\ grows, collisions produce more and more debris. Systems with 
larger mass loss rates evolve more rapidly. Thus, $\alpha$ declines
with \vsqd. 


To construct a simple analytical relation for $\alpha$, we derive 
least-squares fits to the data in Fig.~\ref{fig: alpha1}.  Models with 
$\alpha = \alpha_1 (v^2/\qdstar)^{-e_1} + \alpha_2 (v^2/\qdstar)^{-e_2}$ 
yield $\alpha_1$ = 38.71, $e_1$ = 1.637, $\alpha_2$ = 16.32, and 
$e_2$ = 0.620 (power-law size distribution) and
$\alpha_1$ = 13.00, $e_1$ = 1.237, $\alpha_2$ = 20.90, and $e_2$ = 0.793
(equilibrium size distribution). For the power-law size distribution, the 
model matches the data to better than 5\% over the entire range in \vsqd. 
Although waviness in $\alpha$ for the equilibrium size distribution 
precludes such a good match for all \vsqd, the model agrees within 5\% 
for $ \vsqd \lesssim$ 3000.

To identify a second equation for $\rmax$, we first set the boundary between 
catastrophic and cratering collisions.  We define $f_c$ as the critical ratio
of the collision energy $Q_c$ to the binding energy \qdstar\ which separates
catastrophic and cratering outcomes.  If all particles have the same velocity 
$v$, collisions among more massive particles have larger center-of-mass collision 
energy $Q_c$.  Thus, we can adopt a maximum $x$, \xcc, which results in a 
cratering collision. Collisions with $x > \xcc$ result in catastrophic outcomes.

In principle, establishing \xcc\ is straightforward.  Recalling the mass 
ejected in a collision when $b_d$ = 1, $m_e = 0.5 (m_1 + m_2) Q_c / \qdstar$, 
we require $Q_c / \qdstar < f_c$ for cratering and $Q_c / \qdstar \ge f_c$ for 
catastrophic fragmentation.  Adopting a value for $f_c <$ 1 results in a 
quadratic equation for $\xcc^3$, which has real solutions for 
$\vsqd \ge 8 f_c$ and one solution for $\xcc\ \le$ 1. 

With \xcc\ known, we derive an expression for 
$\dot{r}_{max} = \dot{m}_{max} / 4 \pi \rho \rmax^2 $ = \linebreak
$-\int^{x_{cc}}_{0}~dx~dN_1/dt~(m_e~-~m_2)$:
\begin{equation}
\dot{r}_{max} = -
\left ( {\epsilon r_0 \over 96 t_0 } \right )
\left ( { v^2 \over 4 \qdstar } X_1(x_{cc}) - X_2(x_{cc}) \right )
\end{equation}
where
\begin{equation}
X_1 = \int^{x_{cc}}_{0} { x^{-1/2} (1 + x)^2 dx \over (1 + x^3) } 
= 2~{\rm tan^{-1}}({ \xcc^{1/2} \over x - 1 }) ~
\label{eq: x1}
\end{equation}
and
\begin{equation}
X_2 = \int^{x_{cc}}_{0} x^{-1/2} (1 + x)^2 dx 
= ( 3 \xcc^{1/2} + 10 \xcc\ + 15 ) \left ( { 2 \xcc^{1/2}  \over 15 } \right ) ~
\label{eq: x2}
\end{equation}

Defining
\begin{equation}
\beta = { \epsilon \alpha \over 96 } \left ( {v^2 \over 2 \qdstar} X_1(x_{cc}) - X_2(x_{cc}) \right ) ~ ,
\label{eq: beta}
\end{equation}
we have a simple expression for $\dot{r}_{max}$:
\begin{equation}
\dot{r}_{max} = -\beta { M \over M_0 } { r_0 \over \alpha t_0 } ~ ,
\label{eq: rm1}
\end{equation}
For the standard power-law size distribution 
$N(r) = N_0 r_{max}^{-3.5} x^{-3.5}$,
there is a simple solution to the system of two equations 
(eqs.~\ref{eq: mdisk} and \ref{eq: rm1}) for the two 
unknowns $M$ and \rmax:
\begin{eqnarray}
r_{max}(t) & = & { r_0 \over (1 + t/\tau_0)^{\gamma} }
\\
M_d(t) & = & { M_0 \over (1 + t/\tau_0)^{(1 + \gamma)} }
\end{eqnarray}
where
\begin{equation}
\gamma = \frac{\beta}{1-\beta}
\label{eq: gamma}
\end{equation}
and
\begin{equation}
\tau_0 = (\gamma + 1) t_c = (\gamma + 1) \alpha t_0 ~ .
\label{eq: tau0}
\end{equation}
Using a more general expression for the size distribution -- e.g., 
$N(r) \propto N_0 f(x) r_{max}^{-3.5} x^{-3.5}$ where $f(x)$ is some function 
which relates the standard power-law to the general size distribution --
leads to the same result except for modest changes to the integrals
$X_1$ and $X_2$. Because our main focus is on the time variation
of \rmax\ and $M_d$, we proceed with the solution in eqs.~13--16.

The form of the equations for \rmax\ and \md\ mirror those in the 
standard analytical model. When \rmax\ is constant in time,
$\gamma$ = 0.  At late times, \rmax\ and \md\ follow simple power 
laws: $\rmax(t) = r_0 (t / \tau_0)^{-\gamma}$ and
$\md(t) = M_0 ( t / \tau_0)^{-(\gamma + 1)}$.

Connecting the evolution of \rmax\ and \md\ to the dust luminosity
\ld\ is straightforward. In the standard analytical model, 
$L_d = L_0 / (1 + t / t_0 )$, where $L_0$ depends on the total 
cross-sectional area \ad\ of the swarm of solids. Expressing \ad\ in 
terms of a time-dependent \md\ and \rmax,
\begin{equation}
L_d = { L_0 \over (1 + t / \tau_0)^{-(1 + \gamma/2)} } ~ .
\label{eq: ld1}
\end{equation}
In this expression, the $\gamma / 2$ component results from
the relationship between $L_0$ and \rmax: $L_0 \propto \rmax^{-1/2}$.

Independent of the input parameters, the simple solutions for 
$\rmax(t)$, $M_d(t)$, and $L_d(t)$ yield several robust results.
At early times, the evolution follows standard analytical models
with constant \rmax: $M_d$ and $L_d$ fall to half of their initial
values in one collision time $\alpha t_0$. After several collision
times, \rmax\ starts to approach the asymptotic result,
$\rmax\ \propto t^{-\gamma}$. On the same time scale, $M_d$ and
$L_d$ also begin to follow power-law declines with an exponent
$1 + \gamma$ for $M_d(t)$ and 
$1 + \gamma/2$ for $L_d(t)$.

For any adopted $f_c \le 1$, any initial size distribution, and 
any \vsqd\ $\le$ 4 (\vsqd\ $\ge$ 5), the model predicts the largest 
objects grow (diminish) with time.  Once $f_c$ is known, other 
aspects of the model (including a specific \vsqd\ where 
$\dot{r}_{max}$ = 0) follow uniquely. In practice, however, there 
is no clear boundary between cratering and catastrophic collisions. 
For this study, we use the results of numerical simulations to 
establish $\tau_0$ and $\gamma$.

In addition to $f_c$, the analytic model relies on a constant 
\qdstar\ and the exponents, $b_d$ and $b_l$, in the relations for 
the ejected mass and size of the largest object in the ejecta.
Variations in $b_l$ have modest impact on the evolution of \rmax, 
\md, and \ld; however, small differences in $b_d$ produce measurable
changes in the evolution of \rmax\ and \ld\ \citep{kb2016a}. While
\citet{kb2016a} did not discuss how outcomes with constant 
\qdstar\ differ from those where \qdstar\ varies with $r$, they
note that the evolution of \ld\ in planet formation simulations is
not sensitive to the form of \qdstar\ \citep[see also][]{kb2008,
kb2010,kb2012}. We return to this issue in \S\ref{sec: comps-disc}.

\section{COMPARISON WITH NUMERICAL SIMULATIONS}
\label{sec: comps}

To test the analytical model, we compare with results from numerical
simulations of collisional cascades at 1~AU and at 25~AU. As in
\citet{kb2016a}, we use \orch, an ensemble of computer codes developed
to track the formation and evolution of planetary systems. Within the
coagulation component of \orch, we seed a single annulus with a swarm
of solids having minimum radius \rmin\ and maximum radius \rmax. The
annulus covers 0.9--1.1~AU at 1~AU (22.5--27.5~AU at 25~AU).  At 1 AU
(25~AU), the solids have initial mass $M_d$ = 5~\mearth\ (700~\mearth),
mass density $\rho_s$ = 3~\gcmc\ (1.5~\gcmc), surface density 
$\Sigma_0$ = 106~\gcms\ (24~\gcms), and collision time
$t_0 \simeq 7.51 \times 10^3$~yr ($2.07 \times 10^6$ yr). 

To evolve this system in time, the code derives collision rates and 
outcomes following standard particle-in-a-box algorithms. For these
simulations, the initial size distribution of solids follows a 
power-law, $N \propto r^{-3.5}$, with a mass spacing between mass bins
$\delta \equiv m_{i+1} / m_i$ = 1.05--1.10.  The orbital eccentricity 
$e$ and inclination $i$ of all solids are held fixed throughout the 
evolution: $e_0$ = 0.1 at 1~AU (0.2 at 25~AU) and $i_0$ = $e_0 / 2$. 

In any time step, all changes in particle number for 
$N \le 2 \times 10^9$ are integers. The collision algorithm uses
a random number generator to round fractional collision rates up
or down. This approach creates a realistic `shot noise' in the
collision rates which leads to noticeable fluctuations in \rmax\ and 
\ld\ as a function of time. 

Collision outcomes depend on the ratio $v^2 / \qdstar$. In our 
approach, $v^2$ depends on $a$, $e$, $i$, and the mutual escape
velocity of colliding particles. Although our formalism also includes 
gravitational focusing \citep[][and references therein]{kb2012},
focusing factors are of order unity. For simplicity, we set 
\qdstar\ = constant; varying the constant allows us to evaluate 
how the evolution depends on the initial $v^2 / \qdstar$.  As 
$\rmax$ declines with time, $\vsqd$ also slowly declines. Thus,
we expect some deviations from the predictions of the analytical
model.  For additional details on algorithms in the coagulation 
code, see \citet[][and references therein]{kl1998,kl1999a,kb2001,
kb2002a,kenyon2002,kb2004a,kb2008,kb2012,kb2016a}.

\subsection{Results at 1~AU}
\label{sec: comps-1au}

\begin{figure} 
\includegraphics[width=6.5in]{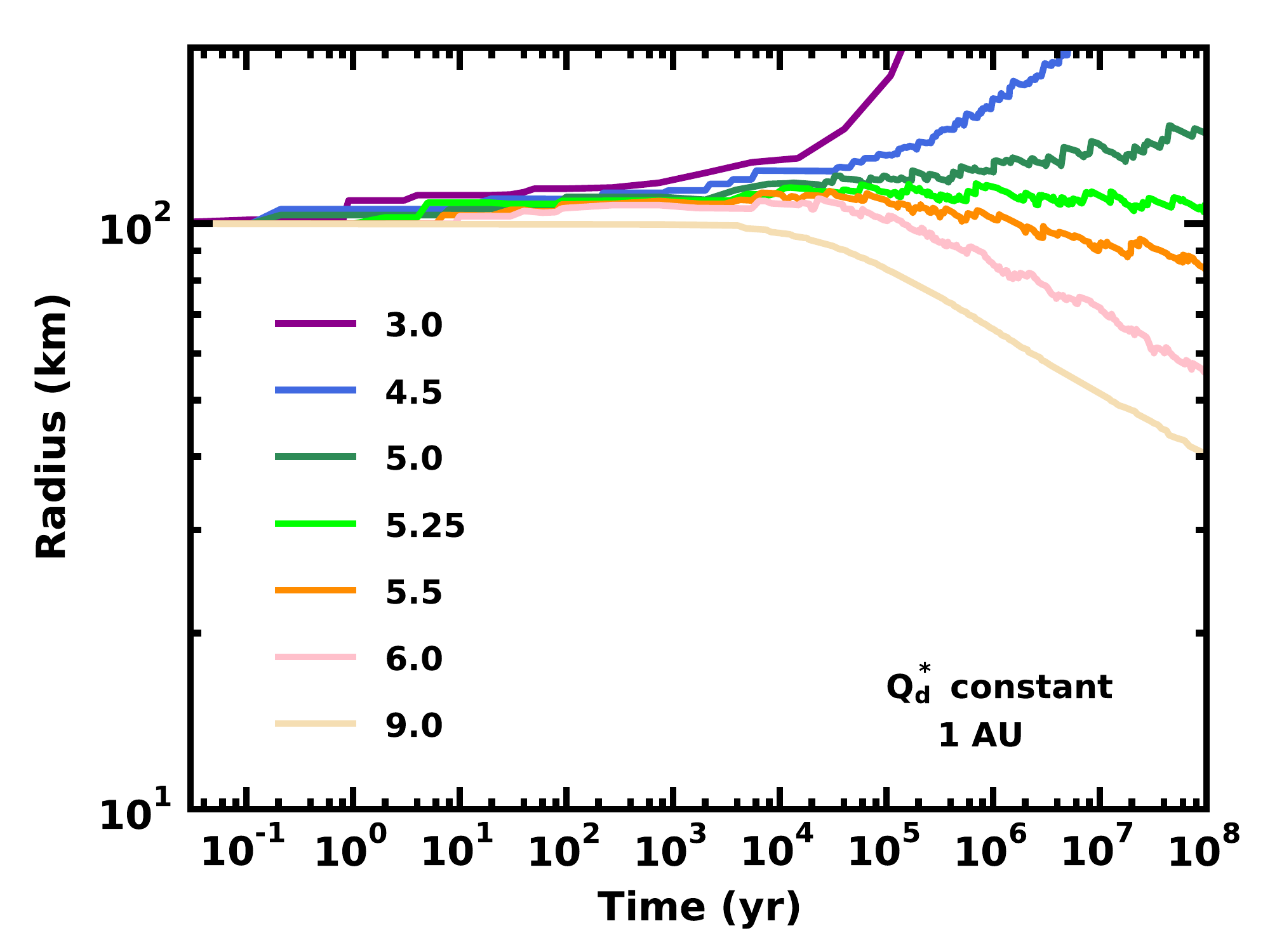}
\vskip 3ex
\caption{
Time evolution of \rmax, the size of the largest object in the cascade, as a 
function of $v_2 / \qdstar$ for numerical simulations of collisional evolution 
at 1~AU.  The legend associates $v^2/\qdstar$ with each curve. In systems with 
$\vsqd \lesssim$ 5.25, the largest object grows with time. 
When $\vsqd \gtrsim$ 5.25, catastrophic and cratering collisions 
steadily remove material from the largest objects.
\label{fig: rmax1}
}
\end{figure}

\begin{figure} 
\includegraphics[width=6.5in]{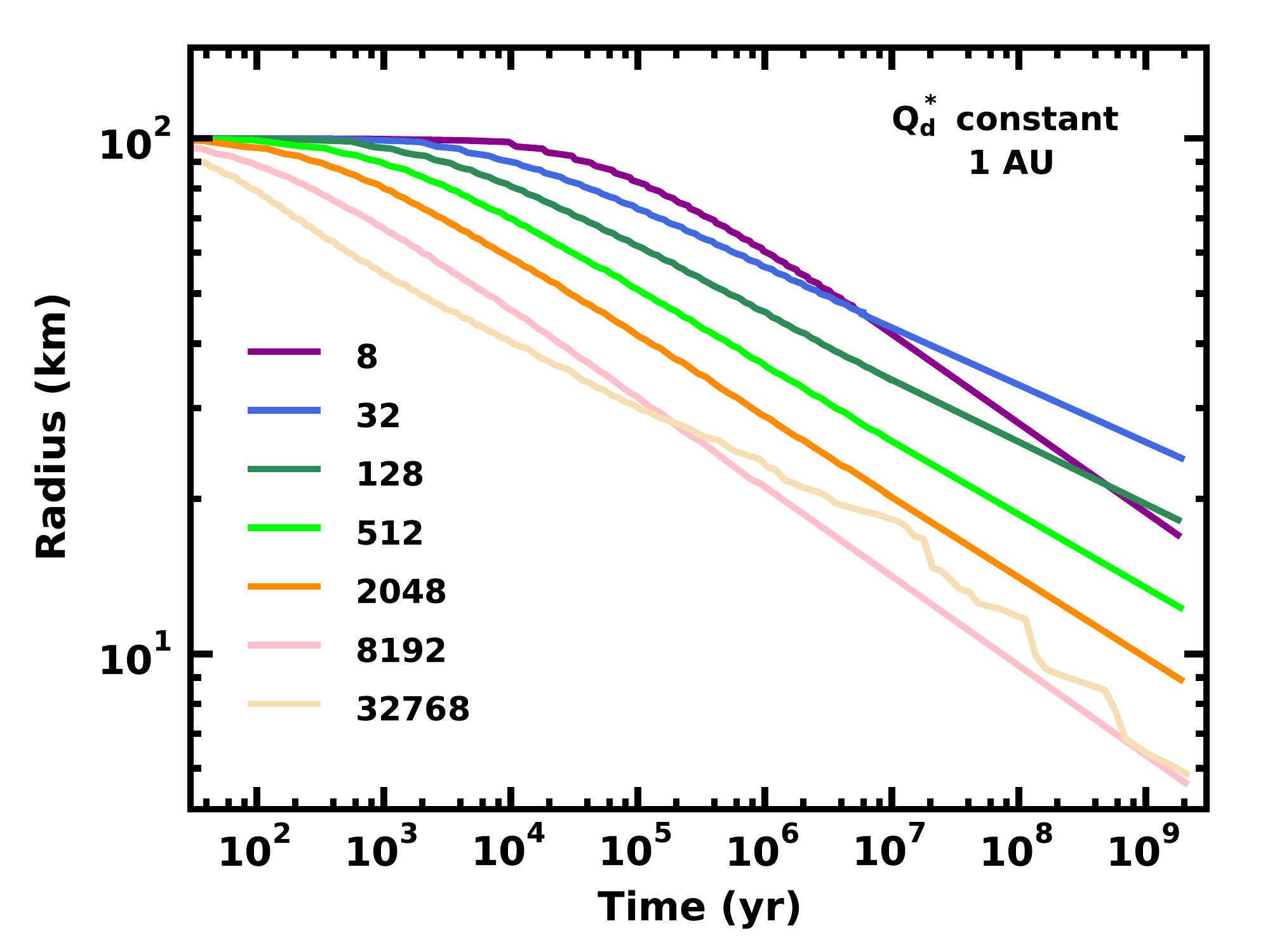}
\vskip 3ex
\caption{
As in Fig.~\ref{fig: rmax1} for calculations with 
$\vsqd \ge$ 8.  Systems with larger $\vsqd$ have more 
destructive collisions and generally evolve more rapidly.
\label{fig: rmax2}
}
\end{figure}

Figs.~\ref{fig: rmax1}--\ref{fig: rmax2} illustrate the evolution 
of the largest objects in a collisional cascade at 1~AU \citep[see 
also][]{kb2016a}.  When $\vsqd \lesssim $ 8 (Fig.~\ref{fig: rmax1}), 
collisions among equal-mass particles yield one larger merged object 
and a substantial amount of debris. Collisions with smaller particles
always produce debris and {\it may} augment the mass of the larger object.

The balance between accretion and mass loss depends on $\vsqd$. For
this suite of simulations where \qdstar\ is independent of particle 
mass density and radius, the largest objects gain (lose) mass when 
$\vsqd \le $ 5.0 ($\vsqd \ge$ 5.5). When $\vsqd\ \approx$ 5.0--5.5, 
growth and destruction roughly balance. Depending on the mix of 
collisions as the system evolves, \rmax\ sporadically increases and 
decreases.  This critical value for $\vsqd$ is close to the value of 
4--5 predicted from the analytical model.

In systems with much larger $\vsqd$ (Fig.~\ref{fig: rmax2}), the
collision time generally decreases monotonically with increasing 
\vsqd.  As predicted by the analytical model, systems with larger 
\vsqd\ initially evolve more rapidly. Once \rmax\ begins to decline, 
however, three evolutionary trends emerge. When \vsqd\ $\approx$ 8--12,
\rmax\ declines rather rapidly. When \vsqd\ $\ge 10^4$ , the initially
rapid evolution in \rmax\ slows considerably and then fluctuates 
dramatically.  At intermediate values (12 $\le \vsqd \le 10^4$), 
\rmax\ evolves much more smoothly at an intermediate rate.

These differences have simple physical explanations. When 
\vsqd\ $\gtrsim 10^4$, the collision parameter 
$\alpha \lesssim 10^{-2}$ (Fig.~\ref{fig: alpha1}). With a
short collision time, $t_c = \alpha t_0 \lesssim 10^3$ yr,
the system loses mass rapidly (see Fig.~\ref{fig: mass1} below). 
Within 1~Myr, the system loses 99.99\% of its initial mass. At
this point, collisions among the largest objects are sporadic;
shot noise dominates the evolution.

When $\vsqd\ \approx$ 8--12, only collisions among roughly equal 
mass objects yield catastrophic outcomes.  Collisions between one 
object and a much smaller particle yield some growth and some debris. 
After several collisions times, systems with $\vsqd \approx$ 8--12 
have (i) relatively more mass in the largest objects and (ii) 
shorter collision times than those systems with $\vsqd \gtrsim$ 12. 
As a result, the largest objects evolve somewhat faster at later 
times when $\vsqd \approx$ 8.

\begin{figure} 
\includegraphics[width=6.5in]{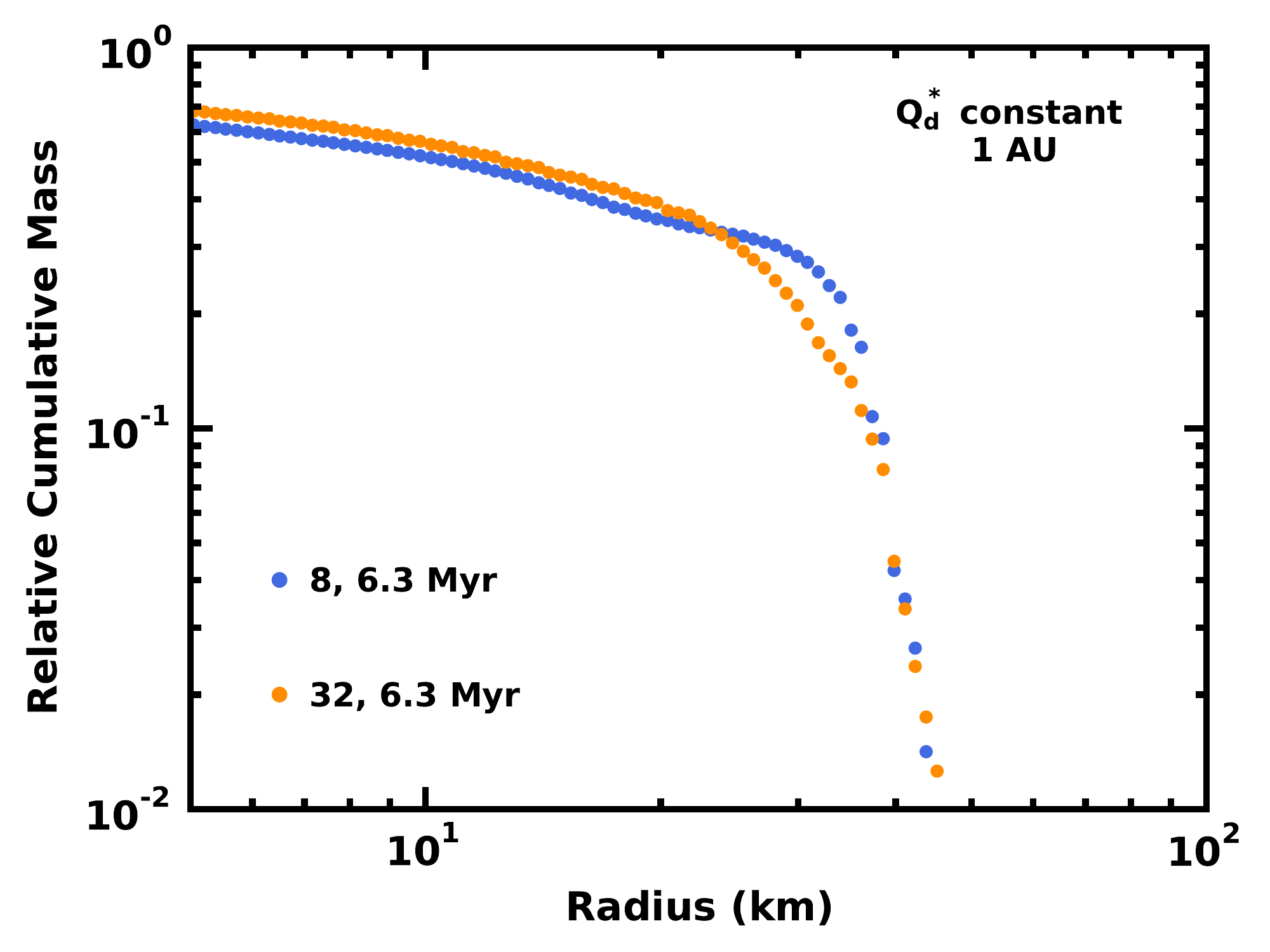}
\vskip 3ex
\caption{
Relative cumulative mass distributions at 6~Myr for calculations 
with \rmax\ = 45~km and either \vsqd\ = 8 (blue points) or 
\vsqd\ = 32 (orange points). Despite identical \rmax, the 
calculation with smaller \vsqd\ has more total mass and more (less) 
mass in objects with $r \gtrsim$ 25~km ($\lesssim$ 25~km). 
The system with \vsqd\ = 8 thus evolves more rapidly than the system
with \vsqd\ = 32.
\label{fig: mass-dist}
}
\end{figure}

To illustrate this point, Fig.~\ref{fig: mass-dist} compares
mass distributions for calculations with \vsqd\ = 8 and 32 at 
6~Myr, when both have the same \rmax. The plot shows the relative
cumulative mass distribution, defined as the cumulative mass from 
\rmax\ to $r$, $M_d(>r)$, relative to the total mass \md\ in the 
grid. This ratio grows from roughly $10^{-2}$ at $r = \rmax$ to 
unity at $r = \rmin$.  For these two calculations, it is clear that 
the system with \vsqd\ = 8 has relatively more mass in solids with 
$r \gtrsim$ 25~km and somewhat less mass in solids with 
$r \lesssim$ 25~km.

In addition to having more mass in large objects, the calculation 
with \vsqd\ = 8 also has more mass overall. Systems with more mass
have shorter collision times (eq.~\ref{eq: t0}). At late times, 
systems with \vsqd\ $\approx$ 8--10 evolve more rapidly than 
systems with \vsqd\ $\approx$ 16--32.

For intermediate \vsqd, the evolution more closely follows 
expectations from the analytical model. Most collisions remove 
mass from the largest objects throughout the evolution. Thus,
these objects gradually diminish in size as the total mass in
the system declines.

Despite differences in the evolution of \rmax, all systems with
a declining \rmax\ lose mass on roughly the collision time scale
$\tau_0 = \alpha (\gamma + 1) t_0$ (Fig.~\ref{fig: mass1}). Although 
there is some shot noise at large \vsqd\ and some growth at small 
\vsqd, the total disk mass always drops smoothly with time. Systems 
with larger \vsqd\ lose mass more rapidly. 

The dust luminosity generally follows the evolution of the total mass 
(Fig.~\ref{fig: lum1}). In every calculation, it takes 10--100~yr for
the size distribution to reach an approximate equilibrium where the
flow of mass from the largest particles to the smallest particles is
similar throughout the grid. Systems with larger \vsqd\ tend to reach
this equilibrium more rapidly and at a somewhat larger \ld\ than systems
with smaller \vsqd. Once this period ends, the luminosity follows a
power-law decline with superimposed spikes in \ld\ due to shot noise.

These results demonstrate that the numerical simulations generally evolve
along the path predicted by the analytical model. After a brief period of
constant \rmax, \md, or \ld, these physical variables follow a power-law 
decline in
time.  To infer the slope of the power-law for each calculation, we perform 
a least-squares fit to $\rmax(t)$, $\md(t)$, and $\ld(t)$. Using an amoeba 
algorithm \citep{press1992}, we derive the parameters $\tau_0$ and $\gamma$ 
from results for $\rmax(t)$ and $\md(t)$. Because our calculations relax to 
an equilibrium size distribution, we add a third parameter $L_0$ to fits for 
$\ld(t)$. Once the fitting algorithm derives these parameters, it is 
straightforward to infer $\alpha$ and $\beta$ using eq.~\ref{eq: t0}, 
eq.~\ref{eq: gamma}, and eq.~\ref{eq: tau0}.

For the complete ensemble of calculations, the amoeba finds each solution in 
20--25 iterations.  Typical errors in the fitting parameters for \rmax\ and
\md\ are $\pm$10\%--20\% in $\tau_0$ and $\pm$ 0.005 in $\gamma$.  Among 
calculations with identical starting conditions, typical variations in 
the fitting parameters are $\pm$5\%--10\% in $\tau_0$ and $\pm0.003$ in 
$\gamma$. Thus, intrinsic fluctuations in $\alpha$ and $\gamma$ are 
comparable to the fitting uncertainties.  Adding the uncertainties in 
quadrature, the errors are $\pm$11\%--22\% in $\alpha$ and $\pm0.006$ 
in $\gamma$. 

Figs.~\ref{fig: rmaxfit1}--\ref{fig: mlfit1} show fits to one set of results
for \vsqd\ = 128. The model $\rmax(t) = r_0 / (1 + t/\tau_0)$ fits the data 
in Fig.~\ref{fig: rmaxfit1} well: the agreement is excellent for 
$t \le 10^4$~yr and $t \ge 10^5$~yr. In between these times, there is a 
small amount of `ringing' as the numerical calculation settles down to 
the standard power-law evolution. For $L_d$ and $M_d$ (Fig.~\ref{fig: mlfit1}),
the agreement between the numerical calculation and the model fits is 
also excellent. 

In this example and all other calculations, the evolution of $M_d$ matches
the model more closely than the evolution of \rmax\ or \ld. As these systems
evolve, changes in \ld\ and \rmax\ consist of a general decline due to the
loss of mass and random fluctuations due to the shot noise inherent in our
collision algorithm. Because larger input \vsqd\ yields shorter collision 
times, these fluctuations grow with increasing \vsqd.  Adopting an appropriate 
measure of these fluctuations enables fits with $\chi^2$ per degree of freedom 
of roughly unity.

For the complete ensemble of calculations, the derived $\alpha$ from fits to
the evolution of \rmax, \md, and \ld\ closely follows predictions for the 
analytical model using the equilibrium size distribution (Fig.~\ref{fig: alpha2}).
Remarkably, independent fits to the evolution of \md\ and \ld\ for the same 
calculation yield nearly identical results for $\alpha$. For the evolution
of \rmax, derived values for $\alpha$ are typically 5\% to 10\% smaller.  
Although this offset is systematic, it is small compared to the uncertainties
in model parameters derived from the amoeba fits. As expected, the analytical 
model provides a poor description of the numerical simulations when 
\vsqd\ $\lesssim$ 8 and growth by mergers is an important process in the 
overall evolution of the swarm.  When $ 10 \lesssim \vsqd \lesssim 10^4$, 
however, the numerical results for $\alpha$ follow the predicted slope very well.

Once $\vsqd \gtrsim 10^4$, the analytical model predicts the numerical results
rather poorly. For these large collision velocities, the evolution of \rmax,
\md, and \ld\ diverge dramatically from each other and from the analytic 
prediction. We associate this divergence with intrinsic shot noise (which grows
as \md\ drops) and the appearance of extreme waviness in the size distribution 
(which causes large fluctuations in the evolution of \rmax, \md, and \ld).

Derived values for $\gamma$ also show clear trends with \vsqd\ 
(Fig.~\ref{fig: gamma1}). As \vsqd\ grows, $\gamma$ declines from 0.15 to
0.1, rises slowly to 0.15, and then fluctuates dramatically. There is a 
modest offset in $\gamma$ for \rmax, \md, and \ld. When 
$10 \lesssim \vsqd\ \lesssim 10^4$, 
$\gamma(\ld) \approx \gamma(\md) + 0.02 \approx \gamma(\rmax) + 0.01$.  
Once $\vsqd \gtrsim 10^4$, $\gamma(\md)  \approx \gamma(\rmax)$;  
$\gamma(\ld)  \approx \gamma(\rmax) + 0.01$. These systematic offsets 
are 2--3 times larger than the uncertainties in $\gamma$ derived from
the amoeba algorithm.

Although the numerical value for $\gamma$ depends on many details, the overall
trends agree with predictions of the analytical model. As \vsqd\ grows, 
collisions are more destructive; the largest objects are diminished more rapidly,
which results in a larger value for $\gamma$. 
Once $\vsqd \gtrsim 10^4$, the extreme waviness in the size distribution sets 
the evolution of \rmax; the analytical model then provides a poor description 
of the system.

For this suite of calculations, the typical $\gamma \approx$ 0.10--0.15 implies
$\beta \approx$ 0.09--0.13. Recalling our definition in eq.~\ref{eq: beta},
the slow variation of $\beta$ as a function of \vsqd\ implies changes in $f_c$ 
with \vsqd.  We infer $f_c \approx$ 1 for \vsqd\ $\approx$ 10, $f_c \approx$ 
0.04 for \vsqd\ $\approx$ 100, $f_c \approx 10^{-3}$ for \vsqd\ $\approx 10^3$,
and $f_c \approx 3 \times 10^{-5}$ for \vsqd\ $\approx 10^4$. The progressive
decline in $f_c$ with increasing \vsqd\ implies a gradual reduction in the 
importance of cratering collisions as the collision energy grows. This result is 
sensible: larger collision energies result in greater frequency of catastrophic
collisions.

\subsection{Results at 25~AU}
\label{sec: comps-25au}

Predictions for the analytical model in \S\ref{sec: theory} are independent 
of $a$.  However, performing a suite of calculations at a different $a$ 
serves several goals:
(i) we make a more robust connection between new calculations and those of 
previous investigators at $a$ = 10--50~AU \citep[e.g.,][]{krivov2005,
krivov2006,lohne2008,gaspar2012a,gaspar2012b}, 
(ii) we develop a better understanding of the impact of the mass 
resolution, stochastic variations, and timestep choices within our code, and
(iii) we infer the impact of changing the particle density $\rho$.
The analytical model is independent of $\rho$ (\S2); however, the numerical 
model uses $\rho$ to calculate the escape velocity of colliding particles, 
which appears in expressions for the gravitational focusing factor and the 
impact velocity.  Although we expect a minor impact on the evolution, 
changing $\rho$ might modify $f_c$ and the mix of cratering and catastrophic 
collisions. 

Aside from a longer collision time, results at 25~AU closely follow those at 1~AU.
In systems with $\vsqd \le$ 5.0 ($\vsqd \ge$ 5.5), large objects gradually gain
(lose) mass with time. For intermediate $\vsqd \approx$ 5.0--5.5, the evolution 
of the largest objects is more chaotic, with mass gain in some periods and mass 
loss during other epochs.  After 10--20~Gyr of evolution with $\vsqd \approx$ 
5.0--5.5, \rmax\ is roughly equal to \r0. Compared to calculations at 1~AU, the
difference in $\rho$ has little influence on the critical \vsqd\ required to 
balance growth and destruction.

For $10 \lesssim \vsqd \lesssim 10^4$, the evolution of \rmax, \md, and \ld\ follow
the analytic model. Calculations with $\vsqd \approx$ 8--10, evolve somewhat more 
rapidly than those with $\vsqd\ \approx$ 20--30, but more slowly than those with 
$\vsqd\ \gtrsim$ 100--200. Once $\vsqd \gtrsim 10^4$, collision rapidly exhaust 
the mass reservoir, leaving the system with few large particles. Shot noise then 
dominates the decline of \rmax.

The analytical model generally fits the evolution of \rmax, \md, and \ld\ extremely 
well.  For $\vsqd \approx$ 5--$10^4$, the amoeba fits derive robust results for the 
fitting parameters $\alpha$, $\gamma$, and $L_0$. In calculations with 
$\vsqd \lesssim$ 5, the largest objects grow with time; shot noise in the growth 
(debris production) rate often leads to poorer fits to the time evolution of 
\rmax\ (\ld). Because \md\ declines in all calculations, the analytical model 
fits the time evolution of \md\ even when \vsqd\ is small. However, the evolution 
of \md\ when $\vsqd \lesssim$ 5 is much slower than the evolution of systems with
larger \vsqd.

Despite substantial differences in the initial mass and a modest change in $\rho$,
calculations at 25~AU yield nearly the same variation of $\alpha$ with \vsqd\ as 
those at 1~AU (Fig.~\ref{fig: alpha3}).  For $\vsqd \gtrsim$ 8--10, results closely 
follow predictions of the analytical model for the equilibrium size distribution.  
Results for the fits to \rmax\ are somewhat closer to these predictions than results 
for fits to \md\ and \ld. However, the differences are fairly negligible compared 
to the uncertainties in amoeba model fits for $\alpha$.

As in the calculations at 1~AU, $\gamma$ clearly correlates with 
\vsqd\ (Fig.~\ref{fig: gamma2}). Although the overall trends in $\gamma$ with
$\vsqd$ are similar at 1~AU and 25~AU, results at 25~AU show a somewhat larger
displacement between the different models. At 25~AU, 
$\gamma(\md) \approx \gamma(\rmax) + g_1$ with $g_1 \approx$ 0.03--0.05 instead of
0.01--0.03. Similarly $\gamma(\ld) \approx \gamma(\rmax) + g_2$ with $g_2 \approx$ 
0.04--0.07 instead of 0.03--0.04.

Calculations at 25~AU also result in somewhat different variations in $f_c$ 
with \vsqd. For \vsqd\ = 10--300, $f_c$ derived at 25~AU tracks results at 
1~AU very closely. When \vsqd\ = 300--$10^4$, $f_c$ is smaller: 
$5 \times 10^{-4}$ at \vsqd\ = 1000 (instead of $10^{-3}$) and
$2 \times 10^{-5}$ at \vsqd\ = 10000 (instead of $3 \times 10^{-5}$).
Compared to the overall change in $f_c$ with \vsqd, these differences
are relatively minor.

\subsection{Discussion}
\label{sec: comps-disc}

The comparisons between results of the numerical simulations and 
expectations from the analytical model are encouraging. Within the
full set of several hundred simulations at 1~AU and at 25~AU, the 
derived evolution of \rmax, \md, and \ld\ matches the predictions 
almost exactly. Repeat calculations with identical starting conditions 
yield nearly identical values for $\alpha$ and $\gamma$. Changing the
particle mass density $\rho$ has minor impact on the results. We 
conclude that the analytical model provides an accurate representation 
of numerical simulations for collisional cascades with a fixed \vsqd.
In the rest of this section, we consider comparisons of our results
with previous studies and discuss how $\gamma$ depends on various
aspects of the calculations.

Previous estimates for the collision time parameter $\alpha$ yield a 
broad range of results. Analytical estimates for \vsqd\ $\gg$ 1 suggest
$\alpha \propto (v^2 / \qdstar)^{-p}$ with $p$ = 5/6 \citep[e.g.,][and references 
therein]{lohne2008,koba2010a,wyatt2011}. Although some numerical 
calculations confirm the analytical result \citep{koba2010a}, others
suggest $p$ = 1.125 \citep{lohne2008} or $p$ = 1 \citep{kb2016a}.

Our analysis clarifies these disparate results. 
For a broad range of \vsqd, we infer 
$\alpha = \alpha_1 (v^2/\qdstar)^{-e_1} + \alpha_2 (v^2/\qdstar)^{-e_2}$ 
with $\alpha_1$ = 38.71, $e_1$ = 1.637, $\alpha_2$ = 16.32, and 
$e_2$ = 0.620 for a power-law size distribution and $\alpha_1$ = 13.00, 
$e_1$ = 1.237, $\alpha_2$ = 20.90, and $e_2$ = 0.793 for the equilibrium 
size distribution.  All previous analytical studies of $\alpha$ for
$\vsqd \gg$ 1 \citep{lohne2008,koba2010a, wyatt2011} agree reasonably well 
with our expectation for the equilibrium size distribution.  In numerical 
simulations, the derived size distribution generally follows the equilibrium 
size distribution \citep{kb2016a}. For the range of \vsqd\ investigated in 
\citet{lohne2008} and \citet{kb2016a} -- \vsqd\ $\lesssim$ 200 -- the predicted 
slope for a single power-law fit to $\alpha$ is 1--1.1, as inferred in these 
two studies. When \vsqd\ is much larger \citep[as in][]{koba2010a}, the 
expected slope is close to 0.8. Thus, various numerical calculations of 
collisional cascades are consistent with one another.

Despite the good general agreement between the analytical model and the 
numerical calculations, there is one clear difference. At late times, the 
analytical model predicts $M_d L_{d}^{-1} r_{max}^{-1/2} \propto t^{-g}$, 
with $g \equiv$ 0. In the numerical calculations, $g \ne$ 0. The non-zero 
$g$ produces offsets in plots of $\gamma(\rmax)$, $\gamma(\md)$, and 
$\gamma(\ld)$ as functions of \vsqd\ (Figs.~\ref{fig: gamma1} and 
\ref{fig: gamma2}). 

Fig.~\ref{fig: allgamma} shows the variation of $g$ with \vsqd. 
Overall, the deviation from the prediction is rather small. Although 
the displacements from zero are somewhat different, the trends 
at 1~AU (blue circles) and at 25~AU (orange circles) are similar: 
(i) a decreasing $g$ at \vsqd\ = 10--100, 
(ii) a roughly constant $g$ at \vsqd\ = 100--$10^4$, and 
(iii) an oscillation at \vsqd $\gtrsim 10^4$. Results at 25~AU are 
somewhat closer to the analytical prediction than results at 1~AU.

The offset of $g$ from zero results from an inability of the numerical
simulations to maintain an equilibrium size distribution. Throughout
every calculation, the derived size distribution is similar but not 
identical to the analytical equilibrium size distribution described
in \S\ref{sec: theory}. As calculations proceed, the numerical size 
distribution also wanders farther away from equilibrium. Relative to
an equilibrium size distribution with \rmin\ = 1~\mum\ and arbitrary 
\rmax, the numerical size distribution usually has somewhat less total 
mass and always has less cross-sectional area. Thus, \md\ and \ld\ decline
faster than \rmax\ relative to the predictions of the analytical model.

There are several possible origins for `non-equilibrium' size 
distributions in our calculations, 
(i) shot noise in the collision rate of the largest objects, 
(ii) non-zero \rmin\ and finite \rmax, and
(iii) finite mass resolution $\delta$ and timestep $\delta t$.
The tests outlined below indicate that differences in $\rho$ have
little influence on the variation of $g$ with \vsqd.

Throughout the course of the evolution, the size distribution is 
the sum of two components: 
(i) an equilibrium piece produced by the steady collisional grinding
of objects with $r \lesssim$ 0.1--0.3~\rmax\ and
(ii) waves of debris generated by occasional collisions among pairs
of the largest objects with $r \gtrsim$ 0.1--0.3~\rmax. Test calculations
demonstrate that steady collisional grinding, without pulses of
debris from collisions of larger objects, yield size distributions
close to the equilibrium size distribution with $g$ almost zero. 
During the pulses, however, the size distribution deviates 
considerably from equilibrium, changing the relationship between 
\md, \ad\ (and thus \ld), and \rmax. Despite the large variations
in the size distribution, $g$ is still fairly close to zero.

In calculations at 25~AU, the larger initial mass reduces shot noise
compared to the calculations at 1~AU. Several calculations with a
factor of ten more mass at either $a$ reduce the absolute value of
$g$. Thus, shot noise is clearly responsible for some of the deviations
of the numerical calculations from the predictions of the analytical 
model.

The non-zero \rmin\ and the finite \rmax\ also contribute significantly
to the non-equilibrium size distribution. For example, when 
\rmin\ = 1~\mum\ ($10^{-3}$~\mum), \rmax\ = 10--100~km, and 
\vsqd\ $\approx$ 100, waviness in the equilibrium size distribution occurs 
for $r \lesssim$ 1--10~cm (10--100~\mum; see Fig.~\ref{fig: equil}).  
Pulses from collisions of 10--100~km objects yield an extra waviness at 
$r \lesssim$ 0.1--1~km. Several test calculations suggest that the 
ability of the numerical calculation to `smooth out' this extra 
waviness depends on the size range between the two sets of waves 
in the size distribution: reducing \rmin\ allows collisional 
processes to reduce the amplitude of the pulse before it reaches 
the intrinsic waviness caused by the non-zero \rmin. In these
calculations, $g$ is closer to zero for \vsqd\ $\lesssim$ 100.

This feature of the numerical calculations explains the trends in
Fig.~\ref{fig: allgamma}. For \vsqd\ $\gtrsim$ 1000, the finite
\rmin\ produces large waves in the equilibrium size distribution 
where shot noise generates pulses in debris production. The 
combination of an intrinsically wavy size distribution at 1--10~km
and wave-like pulses of debris generated from infrequent collisions
of 100~km objects yields a very non-equilibrium size distribution
where the evolution of \md\ and \ld\ are less correlated with the
evolution of \rmax. Thus, $g$ varies rapidly with \vsqd. 

Test calculations suggest that adopting smaller \rmin\ and larger 
initial \rmax\ change the placement of the waves in the relation 
between $g$ and \vsqd\ illustrated in Fig.~\ref{fig: allgamma}. 
Reducing \rmin\ also tends to force $g$ closer to zero; the
change is more dramatic for calculations with \vsqd\ $\lesssim$
100 than for those with \vsqd\ $\gtrsim$ 1000. For these large
values of \vsqd, it is necessary to increase the initial 
\rmax\ significantly to change $g$ dramatically.

Finally, the finite mass resolution and the need for finite time 
steps limit the ability of the coagulation calculations to track 
the analytical model. Figs. 25--26 of \citet{kb2016a} show how
finer mass resolution reduces the noise in numerical calculations
of wavy size distributions.  Although simulations for this paper
with $\delta$ = 1.05--1.10 match analytical predictions very well, 
calculations with smaller $\delta$ would improve the agreement. 
Taking smaller time steps cannot change the impact of a pulse of 
debris on the size distribution; however, smaller steps allow the 
code to smooth out the pulses more evenly.  Calculations with 
smaller $\delta$ and $\delta t$ are very cpu-intensive. Given 
the small differences between the predictions of the analytical 
model and the results of the numerical simulations, more accurate 
calculations are not obviously worthwhile.

For models where \qdstar\ is a function of radius, we expect similar results.
Adopting an expression appropriate for rocky solids at 1~AU, 
$\qdstar$ = $3 \times 10^7 r^{-0.4}$ + $0.3 \rho r^{1.35}$ 
\citep[e.g.,][]{kb2016a}, \vsqd\ $\approx$ 50 for collisions between pairs of
100~km objects. Within a suite of 10 calculations using parameters otherwise
identical to our calculations with constant \qdstar, the variations in 
$\gamma(\rmax)$, $\gamma(\md)$, and $\gamma(\ld)$ are small, 0.01--0.02,
as in calculations with constant \qdstar. Overall, the $\gamma(\rmax)$ values 
are 0.03--0.05 smaller when \qdstar\ is a function of radius. The offsets 
between $\gamma(\md)$, $\gamma(\ld)$ and $\gamma(\rmax)$ are similar, 
$\sim$ 0.02--0.04.

This difference has a simple physical origin. When \qdstar\ is a function of 
radius, \vsqd\ is larger for all solids with $r \le$ 10~km than for larger 
particles. With larger \vsqd, the mass in small particles declines more rapidly 
than the mass in large objects. Calculations with $\qdstar(r)$ then have less 
mass in small particles than those with constant \qdstar\ \citep[e.g., Fig. 15 
of][]{kb2016a}. Compared to a calculation with the same mass in large objects
and constant \qdstar, large objects with $\qdstar(r)$ suffer fewer cratering
collisions and therefore less mass loss; \rmax\ then declines more slowly with
time. Although the overall \ld\ is smaller, it also declines more slowly with
time. Thus, the $\gamma$ factors are somewhat smaller.

Despite the sensitivity of our numerical results to various choices,
applications of the analytical model to 
real data are probably rather insensitive to the choice of $\gamma$ among 
the various possibilities. We suggest setting
$\gamma$ = $\gamma(\rmax) $ = 0.12 for \vsqd\ $\lesssim$ 100--1000 and 
$\gamma$ = $\gamma(\rmax) $ = 0.13 for \vsqd\ $\gtrsim$ 1000. In most real 
systems, the mass of the swarm is rarely large enough to prevent shot noise
from impacting the evolution. The evolution of the cascade then probably 
deviates from the predictions of the analytical model.  In these
circumstances, adopting $\gamma(\md) = \gamma(\rmax) + 0.02$ and 
$\gamma(\ld) = \gamma(\rmax) + 0.03$ should provide an adequate 
representation of the evolution of a real system.

Even though $\gamma$ is small, the evolution of \rmax\ still has an impact on 
the late time evolution of the dust luminosity. After 10--1000 collision times,
systems with a changing \rmax\ are from 15\% to 40\% fainter than those with a
static \rmax. Producing a specific \ld\ late in the evolution therefore requires
a system with a larger initial mass relative to the standard analytical model. 
For some circumstances, the required initial mass is as much as a factor of two 
larger.

\section{SUMMARY}
\label{sec: summary}

We have developed a new analytical model for the evolution of a 
collisional cascade in a ring of solid particles orbiting a massive 
central object. In our derivation for systems with a constant \vsqd, 
\rmax\ the radius of the largest object in the cascade evolves as 
$\rmax = \r0 (1 + t/t_c)^{-\gamma}$, where \r0\ is the initial radius 
of the largest object, $t_c = \alpha (\gamma + 1) t_0$, 
$t_0 = r_0 \rho P / 12 \pi \Sigma_0$, and $\gamma$ is a constant which
depends on $f_c$ the ratio of the collision energy to the critical 
collision energy required for catastrophic collisions.  The mass 
\md\ and the luminosity \ld\ of the solids then evolve as
$\md = M_0 (1 + t/t_c)^{-(\gamma + 1)}$ and
$\ld = L_0 (1 + t/t_c)^{-(\gamma/2 + 1)}$.
The collision time scale parameter $\alpha$ is a simple function of
\vsqd: $\alpha = \alpha_1 (v^2/\qdstar)^{-e_1} + \alpha_2 (v^2/\qdstar)^{-e_2}$ 
with $\alpha_1$ = 13.00, $e_1$ = 1.237, $\alpha_2$ = 20.90, and $e_2$ = 0.793.

The new model applies to cascades in a single annulus of width $\delta a \ge a e$ 
where all particles have the same semimajor axis $a$ and the binding energy of 
solids (\qdstar) is independent of particle size. In disks with a broad range of 
$a$ and constant \rmax, the evolution of \md\ and \ld\ follow more complicated 
functions of time and the inner and outer disk radius \citep{knb2016}. For these
systems, setting $\rmax = \r0 (1 + t/t_c)^{-\gamma}$ as in a single annulus model
and allowing $t_c$ to be a function of $a$ provides a natural extension of the
analytical models discussed here and in \citet{knb2016}. We plan to conduct a set
of numerical calculations to test this idea.

Results from numerical simulations match the analytical model quite well. 
For ensembles of solids at 1~AU and at 25~AU, least-squares fits to the
time evolution of \rmax, \md, and \ld\ yield values for $\alpha$ nearly
identical to model predictions. Although there are minor (0.01--0.02) 
differences in the $\gamma$'s derived from \rmax, \md, and \ld, typical 
solutions require $\gamma \approx$ 0.12--0.13. Thus, the new analytical
model implies somewhat faster declines in total mass and luminosity than
those implied from solutions where \rmax\ is constant in time,
e.g., $\md \propto t^{-1.13}$ instead of $\md \propto t^{-1}$.

The analytical model enables critical tests of coagulation codes for 
planet formation. In our approach, the ability of a coagulation code
to match predictions of the analytical model depends on the spacing 
factor between mass bins $\delta$ and the algorithm for choosing the
time step $\Delta t$. When either $\delta$ or $\Delta t$ is too large, 
it becomes more difficult to match model predictions. Results also
depend on \rmin\ and initial values for \rmax\ and \md. Smaller 
\rmin\ and larger \rmax, \md\ yield better agreement between 
numerical results and analytical predictions.

Along with improved two dimensional models of disks \citep{knb2016}, our
new analytical model should also offer more accurate predictions for the 
long-term evolution of debris disks. In our approach, the dust luminosity 
of a narrow ring declines as $L_d \propto t^{-(\gamma/2 + 1)}$ with 
$\gamma \approx$ 0.15--0.16 instead of the $\gamma \approx$ 0 of standard 
models. The faster decline of the dust luminosity in our models may require 
somewhat more massive configurations of solids than adopted in existing 
studies of debris disk evolution.

We acknowledge a generous allotment of computer time on the NASA
`discover' cluster. Portions of this project were supported by the 
{\it NASA Outer Planets Program} through grant NNX11AM37G. We thank
an anonymous referee for a thoughtful and thorough review that helped 
us to improve the paper. We also thank M. Geller and J. Najita for
cogent comments on several aspects of our approach.

\clearpage

\bibliography{ms.bbl}

\clearpage

\clearpage

\clearpage

\clearpage

\clearpage

\begin{figure} 
\includegraphics[width=6.5in]{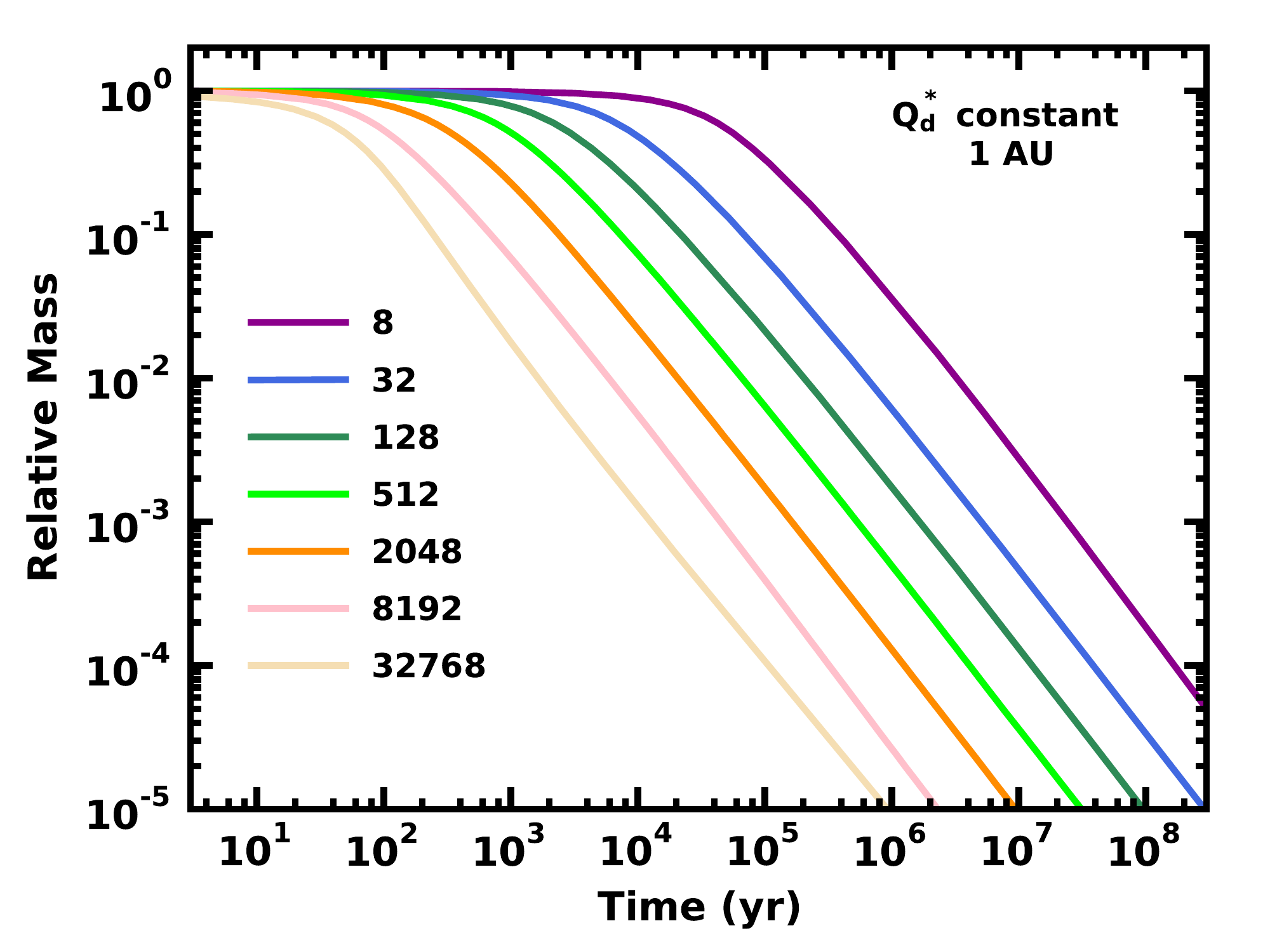}
\vskip 3ex
\caption{
As in Fig.~\ref{fig: rmax2} for the total mass.
\label{fig: mass1}
}
\end{figure}
\clearpage

\begin{figure} 
\includegraphics[width=6.5in]{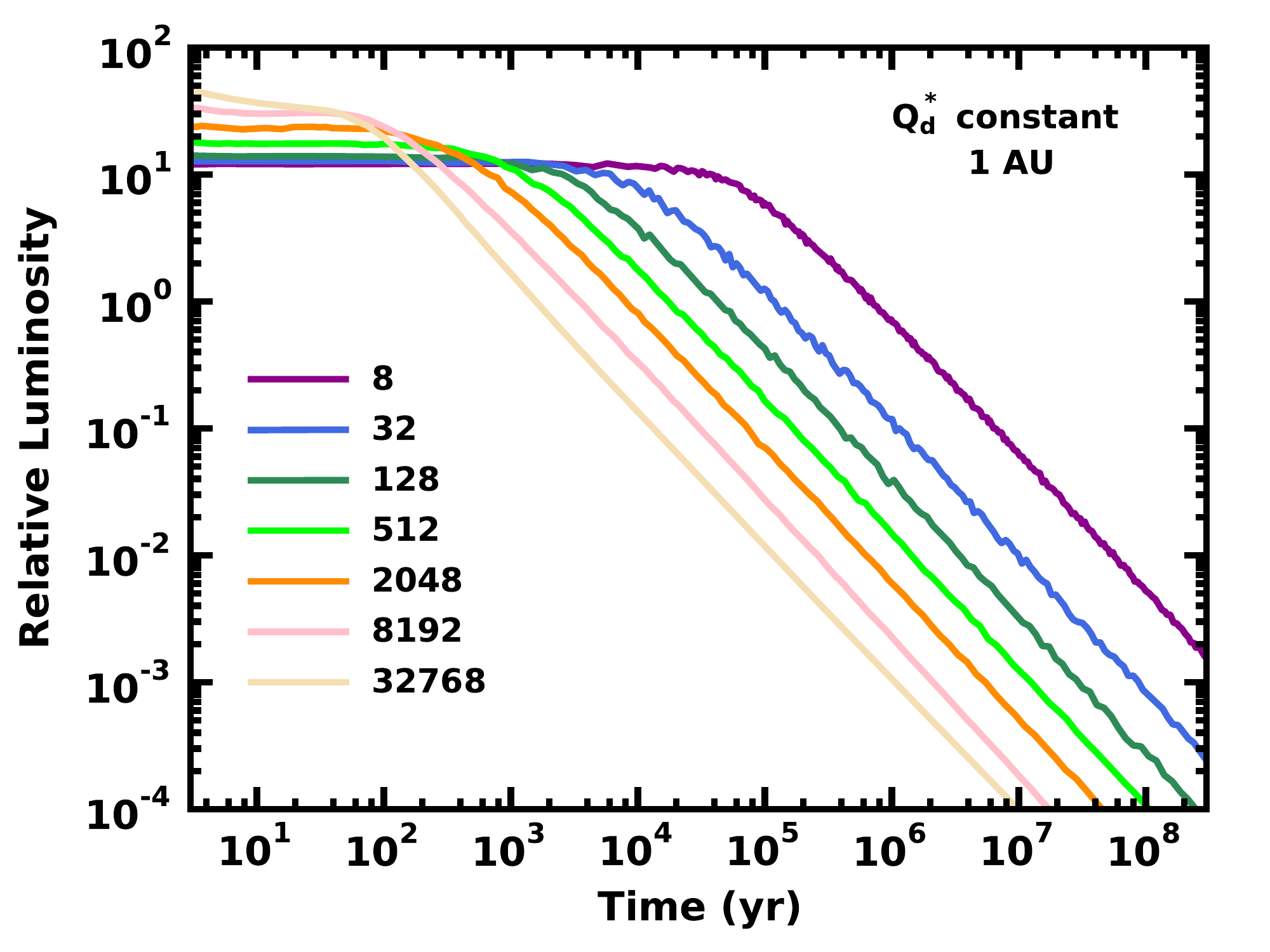}
\vskip 3ex
\caption{
As in Fig.~\ref{fig: rmax2} for the total luminosity.
\label{fig: lum1}
}
\end{figure}
\clearpage

\begin{figure} 
\includegraphics[width=6.5in]{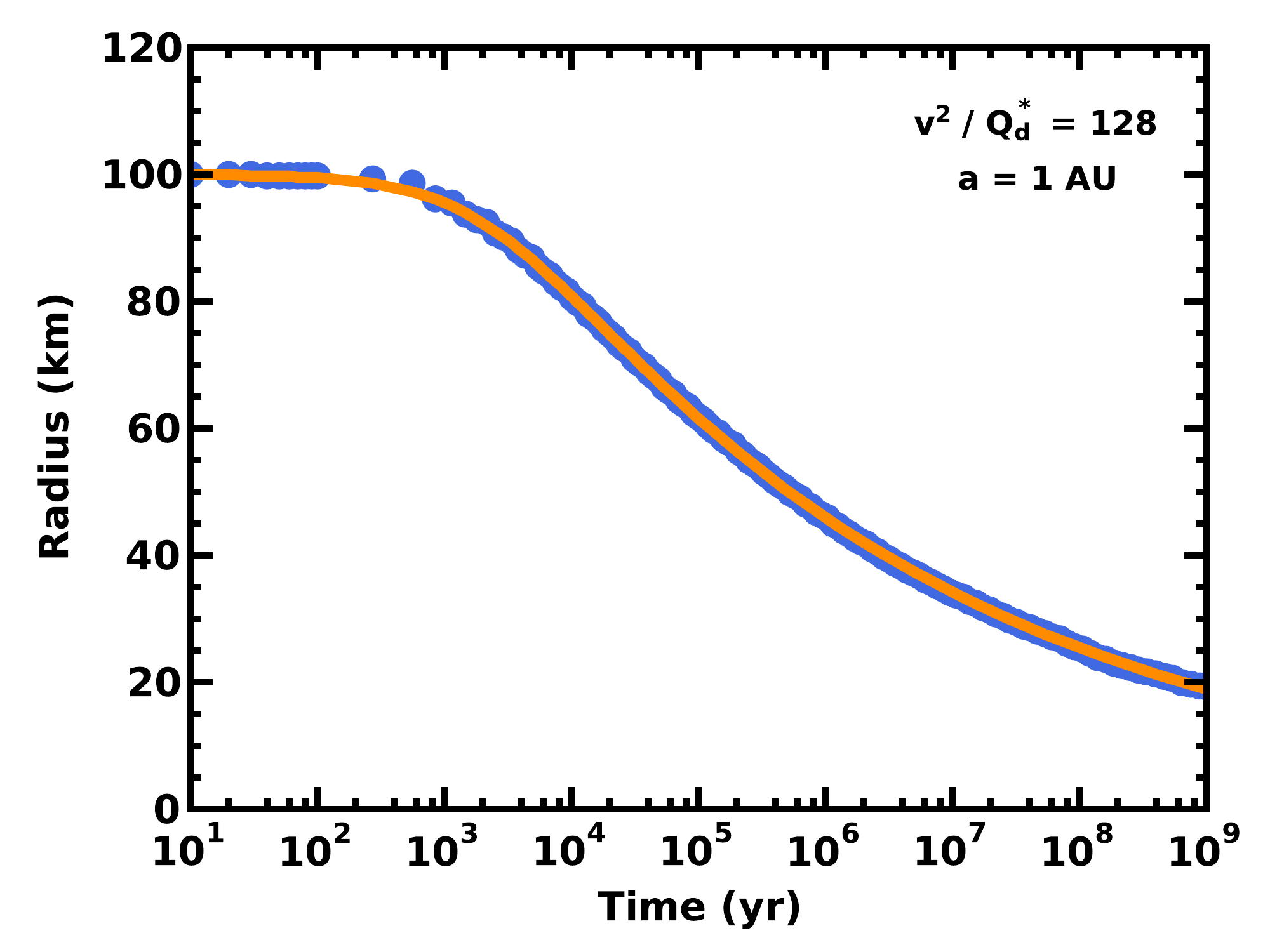}
\vskip 3ex
\caption{
Evolution of \rmax\ in numerical simulations (filled blue circles)
and in the new analytical model (solid orange curve) for a system
with $\qdstar = 2 \times 10^8$ erg~g$^{-1}$ at 1~AU. The analytical 
model matches the numerical calculation to better than 1\% over
$10^1$--$10^9$~yr.
\label{fig: rmaxfit1}
}
\end{figure}
\clearpage

\begin{figure} 
\includegraphics[width=6.5in]{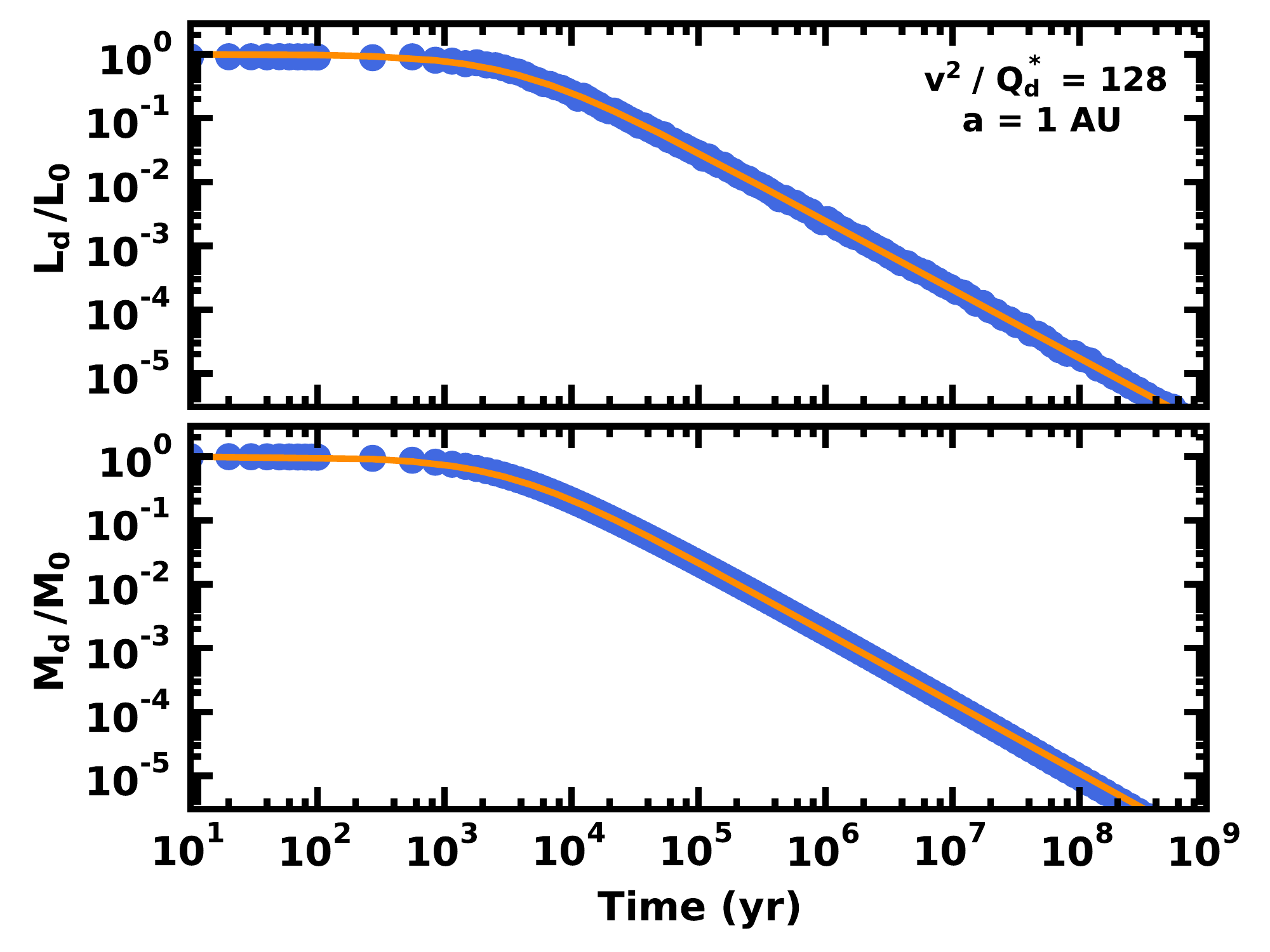}
\vskip 3ex
\caption{
As in Fig.~\ref{fig: rmaxfit1} for the relative disk mass (lower panel)
and the relative disk luminosity (upper panel). The analytical model
matches the numerical calculation to better than 0.5\% in $M_d/M_0$ and
better than 1.5\% in $L_d / L_0$.
\label{fig: mlfit1}
}
\end{figure}
\clearpage

\begin{figure} 
\includegraphics[width=6.5in]{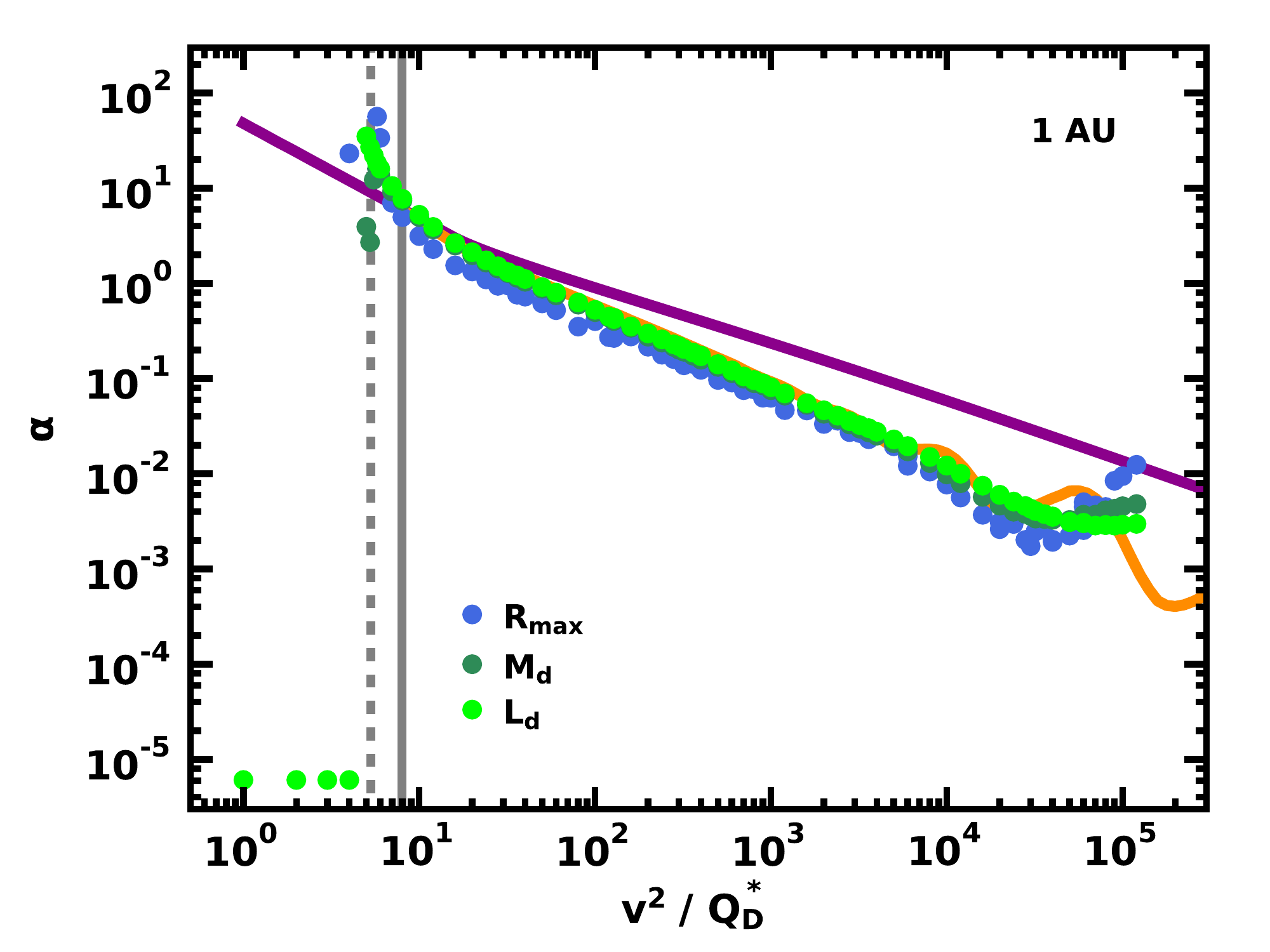}
\vskip 3ex
\caption{
Variation of the derived collision time scale coefficient $\alpha$ 
as a function of $\vsqd$ for collisional cascade calculations at 1~AU. 
Vertical grey lines mark \vsqd\ = 5.25 (dashed) and 8 (solid).
Other solid curves repeat results from the analytical model in
Fig.~\ref{fig: alpha1} for the power-law (magenta) and equilibrium
(orange) size distribution. The numerical results closely follow the
analytical model with the equilibrium size distribution. The legend
associates symbol color with results for \rmax, \md, and \ld
\label{fig: alpha2}
}
\end{figure}
\clearpage

\begin{figure} 
\includegraphics[width=6.5in]{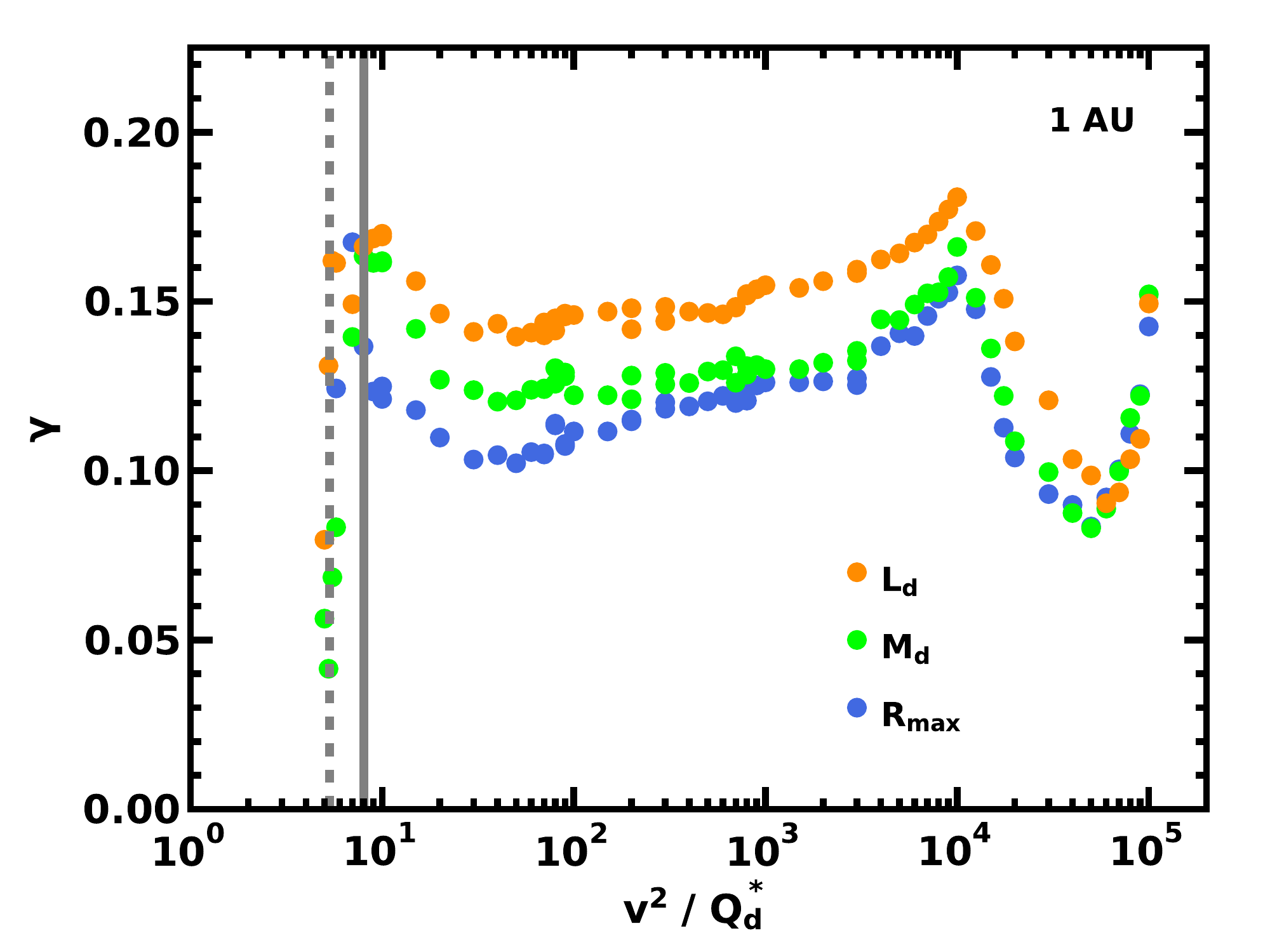}
\vskip 3ex
\caption{
As in Fig.~\ref{fig: alpha2} for the power-law slope $\gamma$ for the
evolution of \rmax. Although the numerical results yield similar values 
of $\gamma$ for the evolution of \rmax\ and \md, the luminosity \ld\ has
a larger $\gamma$ and evolves somewhat more rapidly than expected.
}
\label{fig: gamma1}
\end{figure}
\clearpage

\begin{figure} 
\includegraphics[width=6.5in]{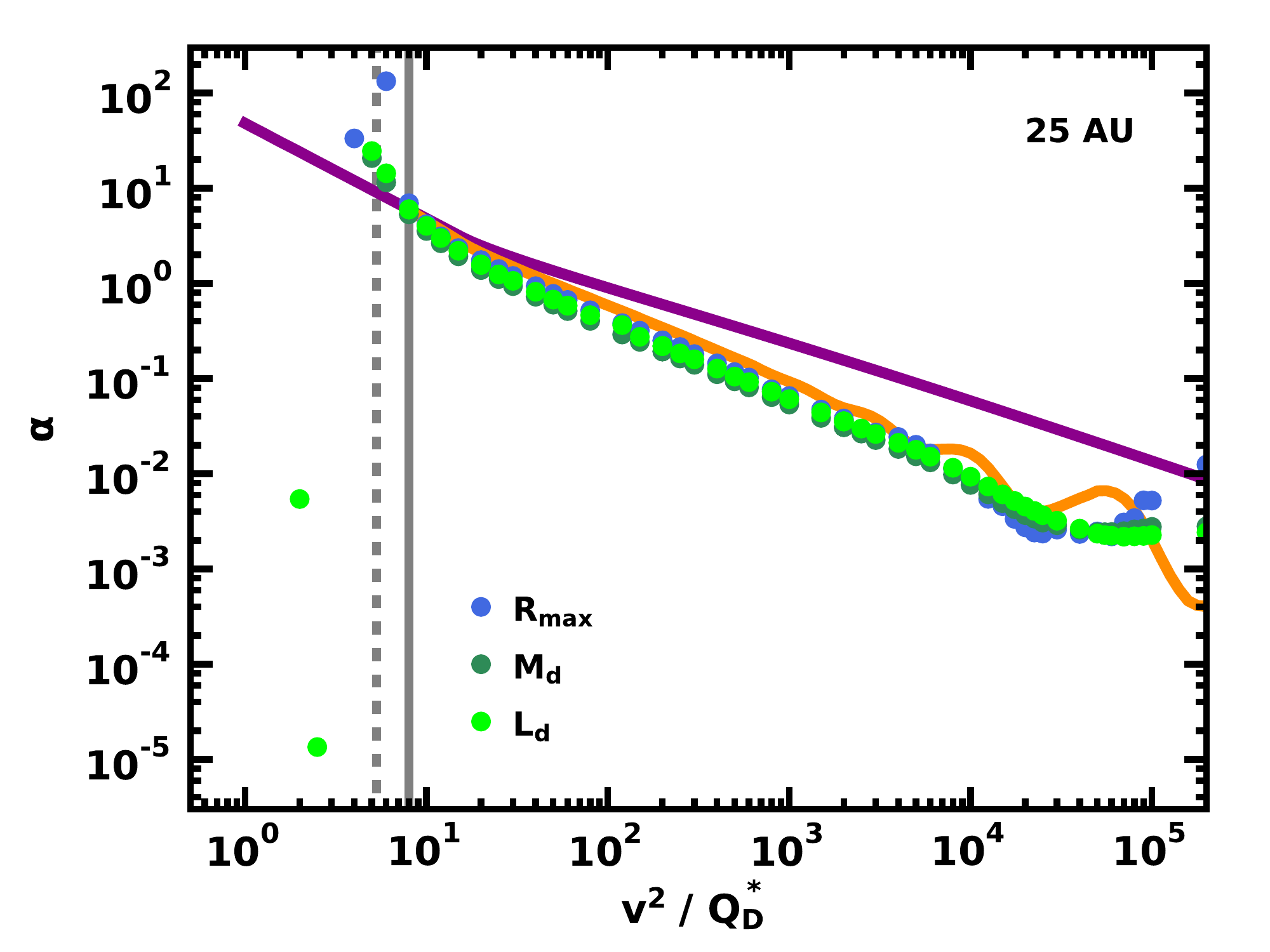}
\vskip 3ex
\caption{
As in Fig.~\ref{fig: alpha2} for calculations at 25~AU.
}
\label{fig: alpha3}
\end{figure}
\clearpage

\begin{figure} 
\includegraphics[width=6.5in]{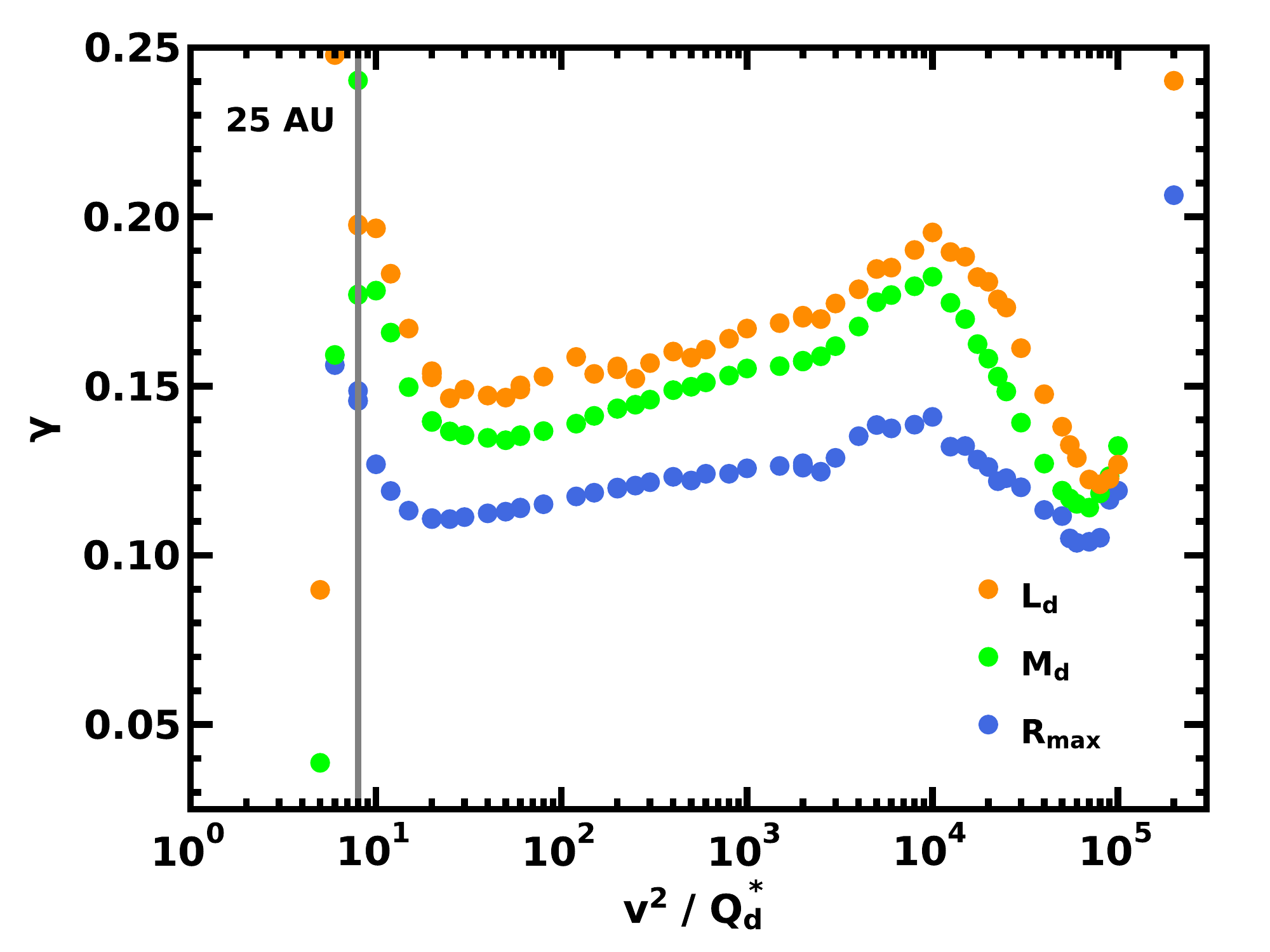}
\vskip 3ex
\caption{
As in Fig.~\ref{fig: gamma1} for calculations at 25~AU.
}
\label{fig: gamma2}
\end{figure}
\clearpage

\begin{figure} 
\includegraphics[width=6.5in]{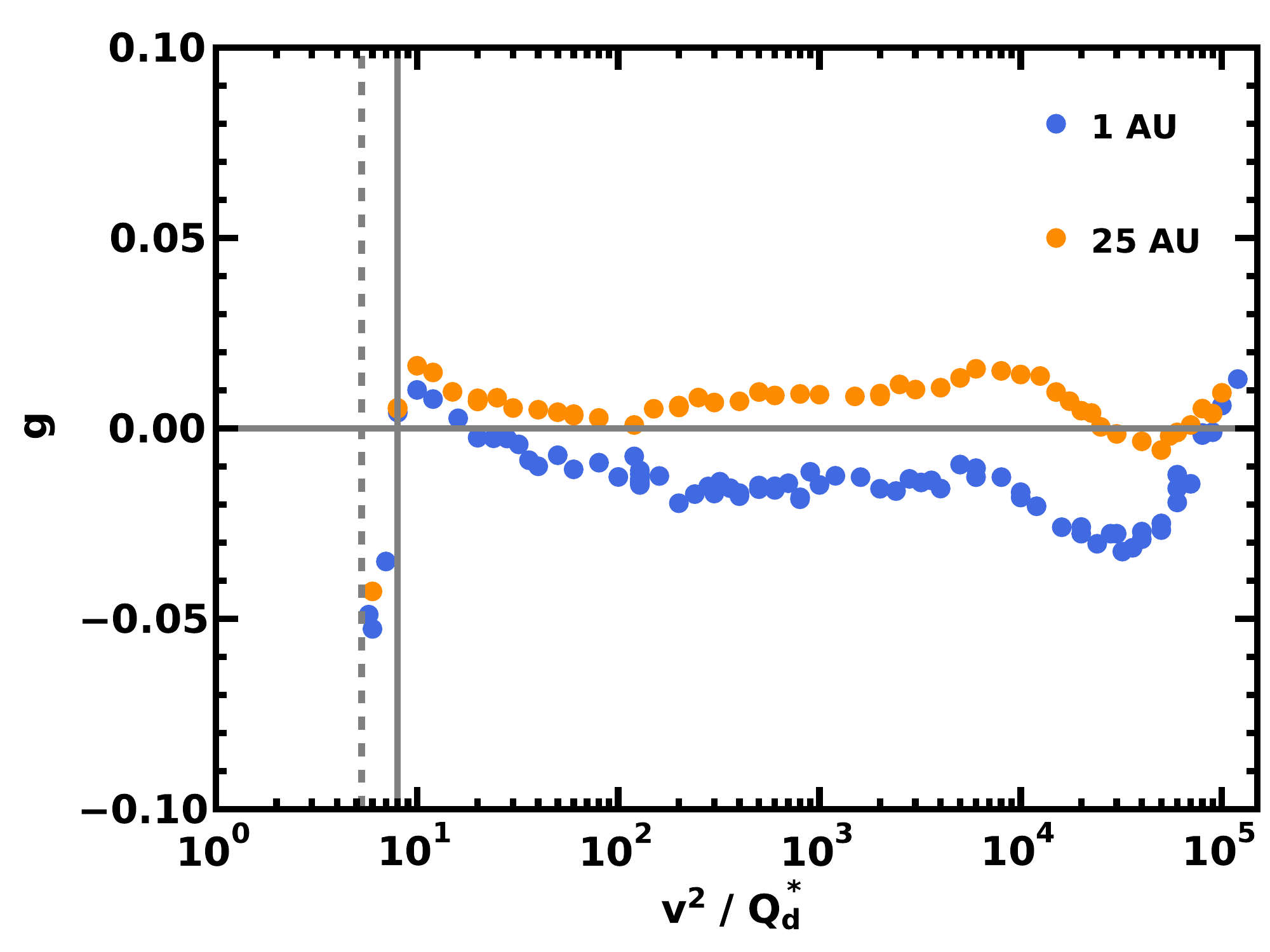}
\vskip 3ex
\caption{
Variation of $g$ = $\gamma(\md) - (\gamma(\ld) + 0.5 \gamma(\rmax))$ 
as a function of \vsqd\ for numerical calculations at 1~AU (blue 
circles) and at 25~AU (orange circles). Vertical grey lines indicate 
the critical values of \vsqd\ for catastrophic disruption of the 
largest objects (solid) and the boundary between growth and destruction 
of the largest objects (dashed). Horizontal solid line indicates the 
predicted $g$ = 0 for the analytical model.  Although the deviations 
between the analytical and numerical model are small, there are clear 
trends with \vsqd.
}
\label{fig: allgamma}
\end{figure}
\clearpage

\end{document}